\documentclass[twocolumn,preprintnumbers,amsmath,amssymb]{revtex4}
\input{epsf}

\usepackage{natbib}
\usepackage[dvips]{graphicx}
\usepackage{mciteplus}

\newcommand{\be}{\begin{equation}}
\newcommand{\ee}{\end{equation}}
\newcommand{\bea}{\begin{eqnarray}}
\newcommand{\eea}{\end{eqnarray}}
\newcommand{\bef}{\begin{figure}}
\newcommand{\eef}{\end{figure}}

\newcommand{\m}{\,{\rm m}}

\newcommand{\simge}{\,{}^>_{\sim}\,}
\newcommand{\simle}{\,{}^<_{\sim}\,}

\def\h#1{$^{#1}$H}
\def\he#1{$^{#1}$He}
\def\li#1{$^{#1}$Li}
\def\be#1{$^{#1}$Be}

\def\eps@scaling{0.96}

\def\showone#1{
  \centering
  \leavevmode
  \epsfxsize=\eps@scaling\linewidth
  \epsfbox{#1.eps}
}

\def\epstwo@scaling{0.48}

\def\showtwo#1#2{
  \centering
  \leavevmode
  \epsfxsize=\epstwo@scaling\linewidth
  \epsfbox{#1.eps} \hfil
  \epsfxsize=\epstwo@scaling\linewidth
  \epsfbox{#2.eps}
}

\begin{document}

\title{Gravitino Dark Matter and the Cosmic Lithium Abundances}
\author{Sean Bailly, Karsten Jedamzik, Gilbert Moultaka} 
\affiliation{Laboratoire de Physique Th\'eorique et Astroparticules, 
UMR5207--CNRS, 
Universit\'e Montpellier II, F--34095 Montpellier, France.}

\begin{abstract}

Supersymmetric extensions of the standard model of particle physics assuming
the gravitino to be the lightest supersymmetric particle (LSP),
and with the next-to-LSP decaying to the gravitino 
during Big Bang nucleosynthesis, are analyzed.
Particular emphasis is laid on their potential to solve 
the "\li7 problem",
an apparent factor $2-4$ overproduction of \li7 in standard Big Bang
nucleosynthesis (BBN), their production of cosmologically important amounts
of \li6, as well as the resulting gravitino dark matter densities 
in these models.
The study includes several improvements compared to prior studies concerning
NLSP hadronic branching ratios, NLSP dark
matter densities, the evaluation of hadronic NLSP decays on BBN,
BBN catalytic effects, as well as updated nuclear reaction rates.
Heavy gravitinos in the constrained minimal
supersymmetric standard model (CMMSM) are reanalyzed, whereas light gravitinos
in gauge-mediated supersymmetry breaking scenarios (GMSB) are studied for
the first time. 
It is confirmed that decays of NLSP staus to heavy gravitinos,
while producing all the dark matter, may at the same time resolve the
\li7 problem. For NLSP decay times $\approx 10^3$sec, such scenarios 
also lead to cosmologically important \li6 (and possibly \be9) abundances.
However, as such scenarios require 
heavy $\simge 1\,$TeV staus they are likely not testable at the LHC. 
It is found that decays of NLSP staus to light gravitinos may lead to
significant \li6 (and \be9) abundances, whereas NLSP neutralinos decaying 
into light gravitinos may solve the \li7 problem. Though both scenarios are
testable at the LHC they may not lead to the production of the bulk of the 
dark matter. A section of the paper outlines 
particle properties required to significantly reduce 
the \li7 abundance, and/or enhance the \li6
(and possibly \be9) abundances, by the decay of an arbitrary 
relic particle.  
\end{abstract}


\maketitle

\section{Introduction}
Supersymmetry is probably the best studied extension 
of the standard model (SM) of particle physics. Softly broken
supersymmetry seems
attractive as it not only may solve the hierarchy problem but 
also lead to gauge coupling unification at the GUT scale and
electroweak symmetry breaking by radiative corrections. 
Additionally, supersymmetry predicts the existence of new particles
whose lightest (LSP), if stable due to some R-parity, may be naturally
produced in the right abundance to provide 
the cosmological dark matter. Here the case of neutralino LSPs has been
widely discussed over at least one-and-half decades, whereas the 
case of gravitino LSPs has received the same widespread
detailed attention only since about five years
\cite{Feng:2003xh, Feng:2003uy, Ellis:2003dn, Feng:2004zu, *Feng:2004mt, 
Roszkowski:2004jd, Baltz:2001rq, Fujii:2002fv, Lemoine:2005hu, Jedamzik:2005ir, Cerdeno:2005eu,
Buchmuller:2006nx, Steffen:2006hw, Kawasaki:2007xb,
Pradler:2007is, Kersten:2007ab, Kawasaki:2008qe}. 
Gravitino dark matter is special due to its superweak interactions
with ordinary matter. It may be produced
at fairly high temperatures $10^5-10^8\,$GeV during reheating after 
inflation and, in case it is the LSP, by the decay of 
the next-to-lightest supersymmetric particle (NLSP) to its 
SM partner and a gravitino. Due to the gravitationally supressed 
interactions of the gravitino the decay with decay time
\begin{eqnarray}
\tau = 
48 \pi \kappa^{-1} \frac{M_{\rm pl}^2 m_{\tilde{G}}^2}{M_{\rm NLSP}^5}
\left( 1- \frac{m_{\tilde{G}}^2}{M_{\rm NLSP}^2}\right)^{-n} \\
\approx 2.4\times 10^4{\rm sec}\,\kappa^{-1}
\biggl(\frac{M_{\rm NLSP}}{\rm 300 GeV}\biggr)^{-5}
\biggl(\frac{m_{\tilde{G}}}{\rm 10 GeV}\biggr)^{2}\nonumber
\label{eq:lifetime}
\end{eqnarray}
typically takes place during or
after Big Bang nucleosynthesis (BBN). 
Here $M_{\rm NLSP}$ and $m_{\tilde{G}}$
are NLSP- and gravitino- mass, respectively, and $M_{\rm pl}$ denotes
the reduced Planck mass, \cite{remark0}.
Gravitino dark matter is therefore subject to constraint,
as it may leave its imprint in the light element abundance yields.

Prediction of the light element abundance yields of \h2 (D), \he3, \he4, and
\li7 by an epoch of Big Bang nucleosynthesis is one of the big successes of
the hot Big Bang model. It has not only lead to the realization that the
Universe has expanded by at least a factor $10^{10}$ in its past, but also
that the bulk of the dark matter must be non-baryonic. In its standard 
version BBN is reduced to a model of one parameter only, the fractional
contribution of the baryonic density to the critical one, $\Omega_bh^2$
(where $h$ is the present
Hubble constant in units of 100 km s$^{-1}$Mpc$^{-1}$).
Five years of observations of the cosmic microwave background 
radiation (CMBR) anisotropies by the WMAP satellite \cite{Hinshaw:2008kr} have lead to the
comparatively accurate determination of 
$\Omega_bh^2 \approx 0.02273\pm 0.00062$ (WMAP-only).
This independent estimate of $\Omega_bh^2$ has promoted
a comparison of observationally inferred primordial light element abundances 
with those predicted in standard BBN to an independent cross-check of the
cosmic SM. Such a comparison is very favorable for D, broadly consistent
yet somewhat inconclusive for \he3 and \he4, and significantly discrepant
for the isotope of \li7. In particular, the primordial \li7/H ratio is
commonly inferred from \li7 absorption lines in the atmospheres of 
low-metallicity halo, or globular cluster stars. Most determinations yield
\li7/H ratios in the range $1-2\times 10^{-10}$ 
\cite{Thorburn:1994ib, Ryan:1999jq, Charbonnel:2005am, 
Asplund:2005yt, Hosford:2008rs}, essentially
the same value as that determined by the Spite's 25 years ago,
$\approx 1.12\pm 0.38\times 10^{-10}$ as for example,
\begin{equation} 
{\rm {}^7Li/H}=(1.23^{+0.68}_{-0.32})\times 10^{-10} \;\;
\end{equation}
\cite{Ryan:1999jq,Hosford:2008rs}, or
\begin{equation}
{\rm {}^7Li/H} =(1.1-1.5)\times 10^{-10} \;\;\nonumber  
\end{equation}
\cite{Asplund:2005yt}.
Some uncertainty remains in the adopted
stellar atmospheric temperatures, which
may lead to somewhat higher estimates,
as in the case of the globular cluster NGC 6397
\begin{equation} 
{\rm {}^7Li/H} =2.19\pm 0.28\times 10^{-10} \nonumber  
\end{equation}
\cite{Bonifacio:1996jv,Bonifacio:2002yx} or the somewhat
controversial result \li7/H$ = (2.34\pm 0.32)\times 10^{-10}$ 
(\cite{Melendez:2004ni}) for
a sample of halo field stars. This is to be compared to the most recent
standard BBN prediction of 
\begin{equation}
{\rm {}^7Li/H} = (5.24^{+0.71}_{-0.67})\times 10^{-10} \nonumber
\end{equation}
\cite{Cyburt:2008kw}
taking into account experimental re-evaluations of the 
for the \li7 abundance most important \he3$(\alpha ,\gamma )$\be7 
reaction rate \cite{NaraSingh:2004vj, Gyurky:2007qq, Confortola:2007nq, Brown:2007sj}. Renewed measurements of this rate has not only increased
the central value for \li7/H by around 25\% , but also reduced 
its error bar considerably, thereby leaving the discrepancy between 
prediction and observation more pronounced 
(around 4.2-5.3$\sigma$ \cite{Cyburt:2008kw}). It is
therefore clear that an additional piece to the puzzle is missing.

Whereas it seems essentially ruled out by now that either revised 
reaction rate data for \he3$(\alpha ,\gamma )$\be7 
or \be7$(D,p)2$\he4 \cite{Coc:2003ce}
or any other reaction,
or a serious underestimate of the stellar atmospheric temperatures
(of order 700K), or even a combination of the two effects may 
resolve the discrepancy, stellar depletion of \li7 stays a viable,
and non-exotic, possibility of resolving the primordial
\li7 problem. Indeed, atmospheric \li7 may be destroyed by nuclear 
burning when transported towards the interiour of the star. This is
also observed in stars with large convective zones, 
but not in those relatively hotter ($T\sim 6000$K) radiation dominated
main-sequence turn-off stars, which are/were believed, after two decades
of research
\cite{Theado:2001xx,*Salaris:2001xx,*Pinsonneault:2001ub,*Boesgaard:2005xe}, 
to preserve at least a large fraction of their 
initial atmospheric \li7. Arguments presented in favor of non-depletion
are (a) the essentially uniform observed \li7 abundance over a wide range 
of different stellar temperatures and metallicities, a pattern which is 
very difficult to achieve when significant depletion was at work, (b) 
the absence of
star-to-star scatter in \li7 abundances expected to arise when, for example,
depletion is induced due to rotational mixing of stars with different
rotational velocities, and (c) the claimed presence of relatively large 
amounts of the much more fragile \li6 isotope in 
some of those stars (cf. \cite{Lambert:2004kn} for a review). 
Thus, taking Occam's razor, many observers believe in the \li7 abundance on
the Spite plateau to be the primordial one. Nevertheless, the situation is
clearly more complicated, since when gravitational settling 
(\lq\lq atomic diffusion\rq\rq ) of heavier elements is included, \li7 is
predicted to be depleted, albeit with a pattern which is not 
observed.
In order to reproduce the observations, another turbulent mixing process of
unknown nature has to be \lq\lq fine-tuned\rq\rq to account 
for the observed pattern \cite{Vauclair:1998it,Richard:2004pj}. 
However, recent observations of the observed
patterns in not only \li7, but also Ca and Ti, in the globular cluster
NGC 6397, lend indeed some support to this idea, though being currently not
statistically overly significant (2-3$\sigma$). 
The combination of atomic diffusion and turbulent mixing has thus been claimed
to be able to account for a \li7 depletion of factor $1.9$ \cite{Korn:2006tv}, a
factor which could go a long way to solve the discrepancy. 

In contrast to \li7, standard BBN (SBBN) yields only negligible production
of \li6/H$\sim 10^{-15}-10^{-14}$ \cite{Nollett:1996ef}. 
This is mainly due to the only \li6 producing reaction in SBBN
D$(\alpha ,\gamma)$\li6 being a quadrupole transition. 
\li6 as observed in higher
metallicity disk stars and in the Sun is believed to be due to cosmic
ray nucleosynthesis, either resulting from spallation reactions
p,$\alpha +$CNO$\to$LiBeB or cosmic ray fusion reactions 
$\alpha+\alpha\to$Li. Since the time-integrated action of standard
supernovae producing cosmic rays is measured by metallicity, it had been
a surprise that three groups independently
confirmed the existence of \li6 in the atmosphere of 
the $[Z]\approx -2.2$ low-metallicity star
HD84937 (\li6/\li7$\approx 0.052\pm 0.019$)
\cite{Smith:1993xx,*Hobbs:1997xx,*Smith:1998xx,*Cayrel:1999xx}, 
a value too high to be comfortably
explained by standard cosmic ray nucleosynthesis. The surprise was even
bigger when the long-term pioneering observational program of 
Asplund {\it et al.} \cite{Asplund:2005yt}
indicated the existence of relatively uniform 
\li6/\li7 ratios $\sim 0.05$ in about ten 
low-metallicity stars, reminiscent of a primordial
plateau (metallicity-independent). Nevertheless, the situation is currently
unclear. The presence of \li6 in these hot stars
has to be inferred by a minute asymmetry
in the atmospheric absorption profile due to the blend of both, the
\li6 and \li7 isotopes. Any individual claimed \li6 detection is 
therefore only at the $2-4\sigma$ statistical significance level. 
Based on observations (of a star which originally, however, was {\it not} 
claimed to have \li6) and complete hydrodynamic 3D non-equilibrium simulations
of stellar atmospheres,
Cayrel {\it et al.} \cite{Cayrel:2007te,*Cayrel:2008hk} 
have recently asserted that line asymmetries 
due to convective motions in the atmospheres could be easily 
misinterpreted as atmospheric \li6, and that the claimed abundances are
therefore spurious. In contrast, Asplund {\it et al.}~\cite{Asplund:2005yt} 
would even infer higher \li6/\li7
ratios, if they were to utilise their own 3D hydrodynamic non-local thermal
equilibrium simulations of line profiles. Further analysis is clearly required.

Concerning cosmic ray production of a putative
\li6/H$\approx 5\times 10^{-12}$, a cosmic ray energy of 100eV per 
interstellar nucleon
is required, whereas standard supernovae generated cosmic rays may provide
only 5 eV per nucleon at such low metallicities 
[Z]$\approx -2.75$ (of the star LP 815-43)\cite{Prantzos:2005mh}.
Such a \li6 abundance requires therefore a very 
non-standard early cosmic ray burst, preferentially acting at higher
redshift \cite{Rollinde:2006zx}, possibly connected to radio-loud quasars and the
excess entropy in clusters of galaxies \cite{Nath:2005ka}, or to a significant 
fraction of baryons entering very massive black holes \cite{Prantzos:2005mh}. 
The requirements are extreme and no compelling candidate has been identified. 
Alternatively, it may be that the \li6 is produced in situ, due to fusion 
reactions in solar flares~\cite{Tatischeff:2006tw}. 

The anomaly in the \li7 abundance and/or the purported one 
in the \li6 abundance, could also be due to physics
operating immediately during or shortly after the BBN epoch, possibly 
connected to the dark matter. It has been long known that \li6 is easily
produced in abundance in non-thermal 
\h3($\alpha$,n)\li6 and \he3($\alpha$,p)\li6 reactions during and after BBN
without much disturbing the other light elements. Here energetic 
\h3 and \he3 may be produced by \he4 spallation- or photodisintegration- 
processes induced by the decay of relic particles
\cite{Dimopoulos:1987fz,Jedamzik:1999di,Jedamzik:2004er,Kawasaki:2004yh}
or residual annihilation of dark matter particles
\cite{Jedamzik:2004ip}. A 
$m_{\chi}\approx 10\,$GeV neutralino of the WMAP density, and with substantial
hadronic s-wave annihilation, as sometimes invoked to explain the DAMA/Libra
anomaly \cite{Bernabei:2008yi}, would, for example, synthesize a \li6 abundance well in excess
of that in HD84937. Similarly, the \li7 abundance may be also affected
by particle decay (but less by annihilation). Early attempts to explain the
\li7 anomaly by \li7 photodisintegration induced by the electromagnetic
decay of a NLSP to a gravitino at $\tau\sim 10^6$sec \cite{Feng:2003uy}, 
were subsequently shown to be in
disaccord with either, a reasonable lower limit on the D/H ratio 
or an upper limit on the \he3/D ratio~\cite{Ellis:2005ii}. However, it
was shown that the injection of 
extra neutrons between the mid and end 
$\sim 100-3000\,$sec of the BBN epoch can 
affect a substantial reduction of the final \li7 
abundance~\cite{Jedamzik:2004er}.
Moreover, if these neutrons are injected energetically, as is the case
during the decay of a weak-scale mass particle, their \he4 spallation may
at the same time be the source of an appreciable \li6 abundance. 

This led to the realization that $m_{\tilde{\tau}}\approx 1\,$TeV 
supersymmetric staus decaying into $m_{\tilde{G}}\approx 100\,$GeV
gravitinos during BBN may resolve two lithium problems at 
once~\cite{Jedamzik:2004er}.
Subsequent more detailed calculations of stau freeze-out abundances,
life times, and hadronic branching ratios 
within the constrained minimal supersymmetric standard model (CMSSM)
\cite{Jedamzik:2005dh} and \cite{Cyburt:2006uv}  confirmed these findings
(cf. also to Ref.~\cite{Kohri:2008cf} and ~\cite{Cumberbatch:2007me}).

The present paper presents a re-/extended analysis of supersymmetric
(SUSY) scenarios with gravitino LSP's which may solve one, or both,
lithium anomalies. The reasons for such a reanalysis are multifold.

\vskip 0.1in
1. Reference \cite{Jedamzik:2005dh} as well as \cite{Cyburt:2006uv}
analyzed scenarios only for large, electroweak scale gravitino masses. 
The present paper studies the case of light $\sim 10\,$MeV-$1\,$GeV 
gravitinos as well. Light gravitinos are typically emerging in 
gauge-mediated SUSY breaking scenarios (GMSB). A consistent
analysis of GMSB scenarios is performed. Finally, for heavier electroweak
scale gravitino masses, as predicted in the CMSSM, an improved scan of
gravitino masses is presented.

\vskip 0.1in
2. The adopted hadronic branching ratios in \cite{Jedamzik:2005dh} 
(based on Ref.~\cite{Feng:2004zu,*Feng:2004mt}) 
are somewhat rough approximations only. For example, as shown in 
\cite{Steffen:2006hw}, the
hadronic branching ratio for the decay of the lighter stau, 
$\tilde{\tau}_1\to\tilde{G}\tau q\bar{q}$, is larger
when production of $q\bar{q}$ pairs by intermediate virtual photons, 
and not only Z-bosons,
is taken into account as well. 

\vskip 0.1in
3. The $q\bar{q}$'s have been noted
to be injected at much lower energy \cite{Steffen:2006hw}
than taken in the approximative treatment of \cite{Jedamzik:2005dh}.
In particular, it has been argued that for accurate BBN calulations
the effective energy $\epsilon_{\rm HAD}$ should be taken as the mean
in the $q\bar{q}$ invariant mass $m_{q\bar{q}}$.

\vskip 0.1in
4. It will be shown in Section II, that taking the
mean invariant mass $m_{q\bar{q}}$ as proposed in \cite{Steffen:2006hw} represents an
approximation as well. Indeed, a fully proper calulation 
requires knowledge of the energy spectrum of the
injected nucleons. 

\vskip 0.1in
5. It has been recently pointed out that charged electroweak mass scale
particles, such as the $\tilde{\tau}$, when present during the
BBN epoch, may lead to interesting catalytic effects 
\cite{Pospelov:2006sc,Kohri:2006cn,Kaplinghat:2006qr}. 
In particular,
catalytic enhancement of \li6 
production~\cite{Pospelov:2006sc,Hamaguchi:2007mp} has an important effect for
stau life times $\tau_x\simge 3000\,$sec, whereas catalytic
enhancement of \li7 destruction~\cite{Bird:2007ge} is less important 
for stau NLSPs. The 
impact of both effects on the $\tau_x\approx 10^3$sec lithium-solving
parameter space has already been investigated in Ref.~\cite{Jedamzik:2007cp}
(see also~\cite{Jittoh:2007fr,Kusakabe:2007fu} 
for other papers on catalysis). The present
paper takes full account of the proposed catalytic processes.

\vskip 0.1in
6. Similarly, Pospelov 
\cite{Pospelov:2007js,Pospelov:2008ta} has made the interesting suggestion that
\be9 may be produced due to catalytic effects. 
These effects are included in the present study and their significance 
is adressed.

\vskip 0.1in
7. Finally, the re-evaluation of 
the main \li7 producing rate
~\cite{Cyburt:2008kw,NaraSingh:2004vj,Gyurky:2007qq,Confortola:2007nq,Brown:2007sj}, as well as a shift
in $\Omega_bh^2$ \cite{Hinshaw:2008kr}, as indicated in the introduction,
has increased the predicted standard BBN \li7 abundance. These
changes are included in the present paper 
as well as recent changes in some catalytic
rections rates \cite{Kamimura:2008fx}.

\section{Dependence of results on nucleon spectrum}

\begin{figure}
\includegraphics[width=.5\textwidth]{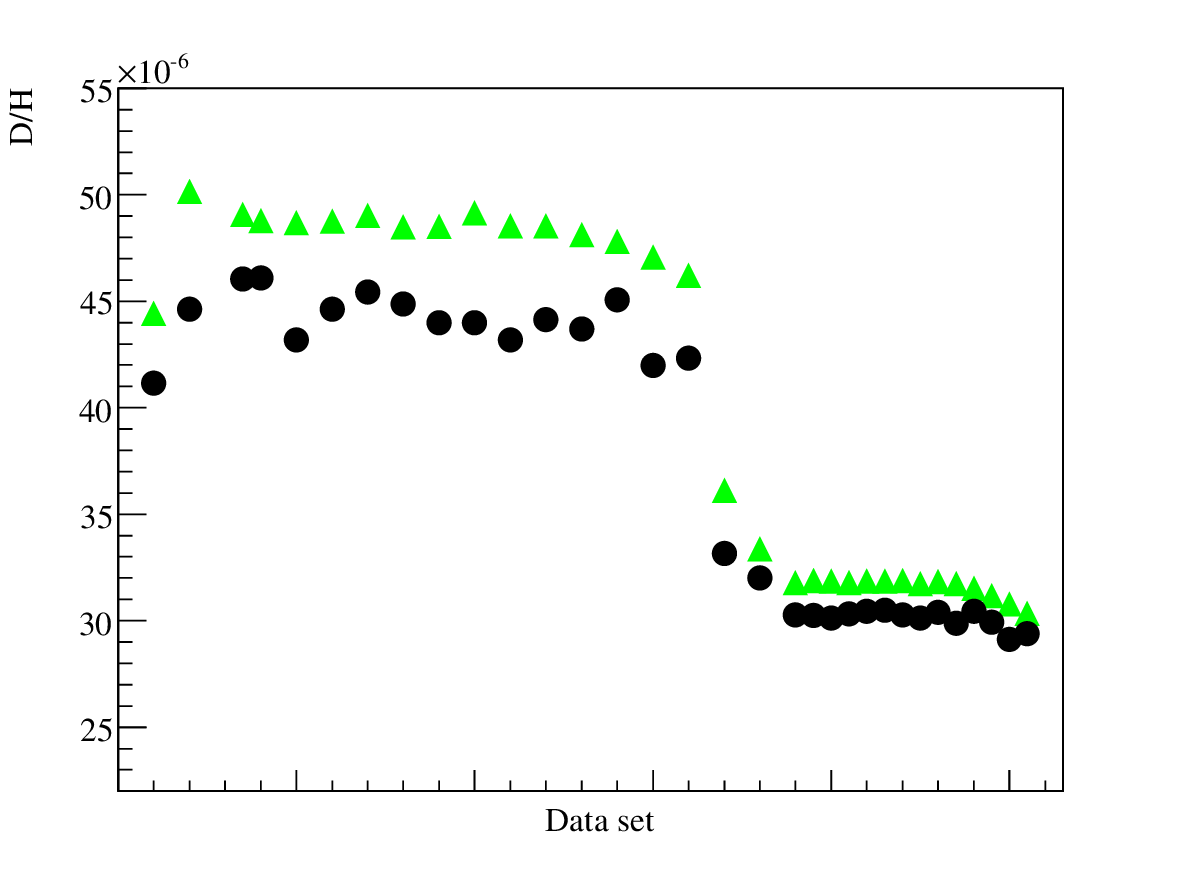}
\includegraphics[width=.5\textwidth]{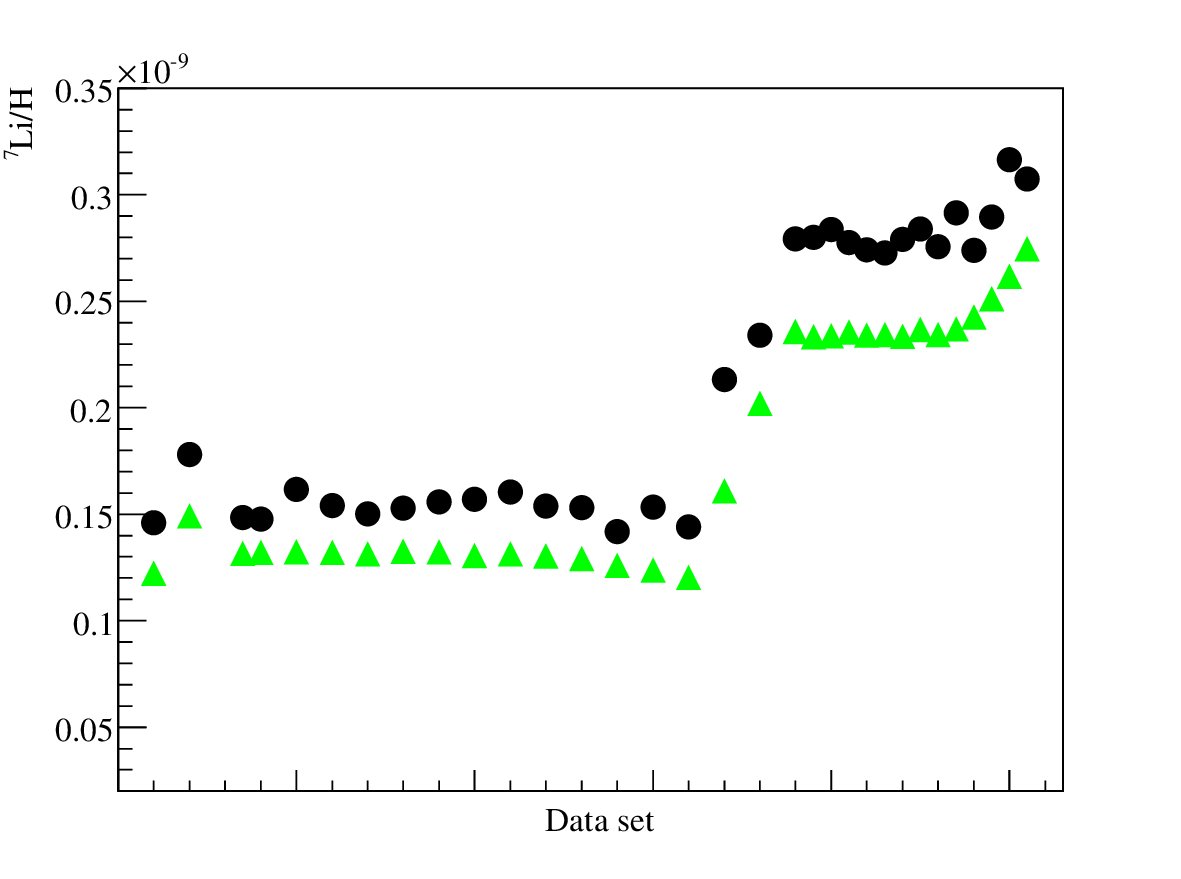}
\caption{D/H (upper) and \li7/H (lower) abundances computed for a 
number of selected models with $m_{\tilde{\tau}}\approx 1\,$TeV and
$\Omega_{\tilde{G}}h^2\approx 0.1$ (see text for further detail) when
(a) the invariant mass approximation is used (green triangles) and
(b) the analysis is performed using a realistic nucleon energy 
spectrum (black dots). The left (right) plateau show models 
with approximate decay times 
$\tau\approx 10^3$sec ($\tau\approx 300-600\,$sec), respectively. They 
correspond to models in the blue (red) regions of 
Fig.~\ref{fig:Omegatau_heavy}, respectively.} 
 \label{fig:points}
\end{figure}

In \cite{Steffen:2006hw} it was argued that the quantity
\begin{equation}
\langle m_{q\bar{q}}\rangle  = 
\frac{1}{\Gamma_{\mbox{\scriptsize tot}}(\tilde{\tau})}  
\int_{m_{q\bar{q}}^{\mbox{\scriptsize cut}}}^{m_{\tilde{\tau}}-m_{\tilde{G}}-m_\tau} 
\!\!\!\!\!\!\!\!\!\!\!\! dm_{q\bar{q}} \, m_{q\bar{q}} \,  \frac{d\Gamma(\tilde{\tau}\rightarrow \tau\tilde{G}q\bar{q})}{dm_{q\bar{q}}}
\label{eq:SteffenEHad}
\end{equation}
should be taken as the \lq\lq effective hadronic energy\rq\rq\ 
$E_{\mbox{\scriptsize Had}}(\tilde{\tau}\rightarrow 
\tau \tilde{G} q \bar{q})$ for the
computation of BBN yields with decaying particles. It was noted that
this is considerably lower for hadronic decays of $\sim $TeV stau's
than the adopted $(m_{\tilde{\tau}}-m_{\tilde{G}})/3$ in \cite{Jedamzik:2005dh}.
Though this is a good comment, 
the situation is much more complicated. In contrast to
electromagnetic decays, there is {\it no} obvious sensible definition 
of an \lq\lq effective hadronic energy\rq\rq\ . 
For example, ten $q\bar{q}$ fluxtubes of $E_{CM} = 100\,$GeV produce about
$13.5$ neutrons and protons (of varying energy), whereas one fluxtube of
$E_{CM} = 1\,$TeV produces only $3.2$ nucleons. Similarly, the injection of
ten $10\,$GeV neutrons at temperature $T = 30\,$keV lead to the
production of $4.3$ thermalized D, 14.4 neutrons, and $4.0$ 
nucleon number $A=3$-nuclei. This needs to be compared to $0.5$ D,
1.45 neutrons and $0.7$ $A=3$-nuclei for one injected 
$100\,$GeV neutron. In both examples one started with
the same initial \lq\lq hadronic energy\rq\rq\ , with the result being 
vastly different. 
The number $\Delta N_i^{casc}$ (neutrons, D, \h3, \li6, ...) of thermalized
nucleons/nuclei $i$ produced per $X$ particle decay at temperature $T$ may
be written in the following form 
\begin{equation}
\Delta N_i^{casc} = B_h\sum_{j=p,n}
\int {\rm d}E_jP^{had}_j(E_j)\frac{{\rm d}N_i^{casc}}{{\rm d}N_j}(E_j,T)
\end{equation}
where $P^{had}_j(E_j) = ({\rm d} N_j/{\rm d}E_j)$ is the
distribution function of nucleons $j$ (protons, neutrons)
with initial energy $E_j$
injected into the plasma by the $X$ decay, normalized such that
\begin{equation}
\int {\rm d}E_j P^{had}_j(E_j) = \langle N_j\rangle
\end{equation} 
gives the mean number of injected nucleons $j$ per hadronic decay.
In the above the quantity ${{\rm d}N_i^{casc}}/{{\rm d}N_j}(E_j,T)$ gives
the produced (and thermalized) nucleons/nuclei $i$ per injected 
nucleon $j$ injected at initial energy $E_j$, 
and $B_h$ is the hadronic branching ratio.
For the computation of $\Delta N_i^{casc}$ a detailed analysis of the
thermalisation of an injected nucleon $j$, as well as the thermalisation
of all produced secondary (and higher) generations of non-thermal nucleons
and nuclei due to energetic nucleon inelastic- and spallation- scattering 
processes 
on thermal nucleons and nuclei is required (cf.~\cite{Jedamzik:2006xz}).

The invariant  mass $m_{q\bar{q}}$ gives the energy $E_{CM}$
of the $q\bar{q}$ fluxtube in the $q\bar{q}$ center-of-mass
reference system. This quantity does {\it not}, however, 
give the energy of the fluxtube in the 
cosmic rest frame (stau rest frame). The latter is much larger.
Nevertheless, approximating $P^{had}_j(E_j)$ by the nucleon energy
distribution in the center-of-mass frame of $q\bar{q}$, resulting after fragmentation
of a $q\bar{q}$ fluxtube of energy 
$\langle m_{q\bar{q}}\rangle$,
is still somewhat useful, since in certain temperature ranges,
such as $50{\rm keV}\simle T\simle 80\,$keV, the BBN yields mostly
only depend on the number of nucleons injected, as long as they are
somewhat energetic $E_{kin}\simge 1\,$GeV. Whereas the energy of
the created nucleons is not an invariant of the reference system,
the number is. However, even this corresponds to an approximation.

For a limited number of scenarios we have computed a realistic
nucleon spectrum. 
This spectrum was then used  as input to perform accurate BBN calculations. 
The scenarios studied are $\sim 1\,$TeV
staus decaying between $\tau\approx 300-1200\,$sec into gravitinos,
with the decay producing $\Omega h^2\approx 0.1$ in gravitinos.
Fig.~\ref{fig:points} shows the yields of D/H and \li7/H  as obtained either
from the realistic nucleon spectrum or from the 
approximation of a nucleon distribution resulting from a $q\bar{q}$ pair of 
energy $\langle m_{q\bar{q}}\rangle$ in its center-of-mass frame.
It is seen that the realistic spectrum gives larger \li7 and lower Deuterium
abundances by $\simle 10-15\%$ as compared to the  
$\langle m_{q\bar{q}}\rangle$ approximation. It is
cautioned however, that the relative error is only small for comparatively
early decays. In what follows, when not stated explicitly
otherwise, all figures shown employ this approximation.

All BBN calculations assume $\Omega_bh^2 = 0.02273$ and employ the
recent reevaluation of the \he3$(\alpha ,\gamma )$\be7 rate by 
\cite{Cyburt:2008kw}. For details concerning the BBN calculations the reader
is referred to \cite{Jedamzik:2006xz} 
(and~\cite{Jedamzik:2007cp}). Catalytic rates are taken from the 
recent papers \cite{Hamaguchi:2007mp} and \cite{Kamimura:2008fx}.
Details relevant to the CMSSM and
GMSB supersymmetric models, NLSP freeze-out abundances, as well as 
the calculation of hadronic branching ratios will be presented elsewhere. 
This analysis was performed using several public numerical codes: 
SuSpect \cite{Djouadi:2002ze}
(a supersymmetric particle spectrum calculator), 
micrOMEGAs \cite{Belanger:2006is} (a
dark matter relic density calculator), 
CalcHEP \cite{Pukhov:2004ca} (an automatic matrix
element generator), and 
PYTHIA \cite{Sjostrand:2006za} (a Monte-Carlo high-energy-physics
event generator).

\section{A Guide to the Model Builder}

\begin{figure}
\includegraphics[width=.45\textwidth]{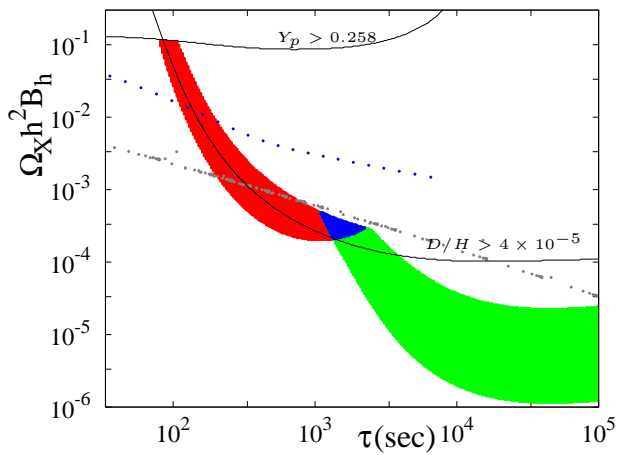}
\caption{
Parameter space in the $\Omega_Xh^2B_h$ versus $\tau$ plane where
decaying relic neutral particles have impact on the \li7 and \li6 abundancies
following Ref. \cite{Jedamzik:2004er}.
Here $\Omega_X$ is the fractional contribution to the critical density the 
decaying particle would have today if it wouldn't have decayed, $h$ is the Hubble constant
in units of $100\rm km s^{-1}Mpc^{-1}$, $B_h$ is the hadronic branching
ratio, and $\tau$ is the particle life time. 
The color coding of the areas
is as follows: (red) \li7/H$< 2.5\times 10^{-10}$;
(green) $0.015<$\li6/\li7$<0.3$; and (blue) both \li7/H$< 2.5\times 10^{-10}$
{\it and} $0.015<$\li6/\li7$<0.3$. Other (relevant) constraints on
light elements are taken to be 
D/H$<5.3\times 10^{-5}$ 
and the helium mass fraction $Y_p < 0.258$, as indicated by a line. 
The black solid line labeled D/H$>4\times 10^{-5}$
shows how lithium-friendly parameter space is reduced when a less 
conservative limit on D/H is applied. 
The figure also shows by the grey and blue
points, following essentially lines, the prediction for stau-NLSPs
in the CMSSM assuming a gravitino mass of $m_{\tilde{G}} = 50\,$GeV
and the prediction for neutralino NLSPs in a GMSB scenario assuming a
gravitino mass of $m_{\tilde{G}} = 100\,$ MeV, respectively.
To produce this figure, the hadronic decay of a $1\,$TeV particle into a
quark-antiquark pair has been assumed. Results for other initial states
and particle masses $M_X$ may vary by a factor of a few such
that the figure should be interpreted as indicative only.}
\label{fig:banana1}
\end{figure}

\begin{figure}
\includegraphics[width=.45\textwidth]{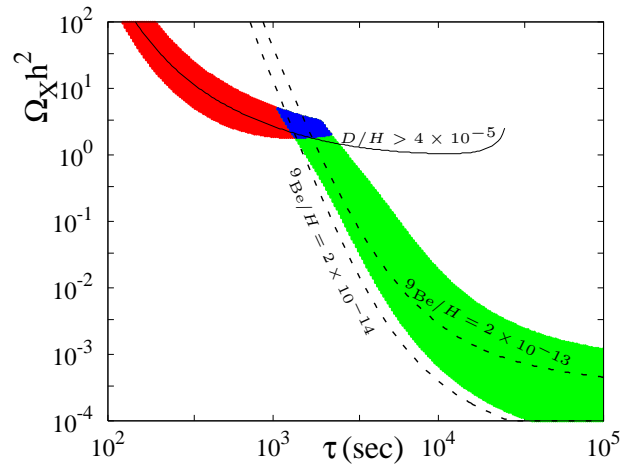}
\caption{
Parameter space in the $\Omega_Xh^2$-$\tau_X$ plane where
{\it charged} relic decaying particles have impact on the cosmic
\li7, \li6, and (possibly) \be9 
abundances without violating other light-element
constraints.  
Here $M_x = 1\,$TeV, $B_h = 10^{-4}$ and hadronic primaries of
1 TeV $q\bar{q}$ fluxtubes have been assumed.
The color code 
is as described in caption Fig.~\ref{fig:banana1}. 
The two dashed lines delimit the band 
$2\times 10^{-14}\simle$\be9/H$\simle 2\times 10^{-13}$,
whereas the solid line indicates D/H$= 4\times 10^{-5}$.}
\label{fig:banana2}
\end{figure}

In this section we discuss, in a generic way, 
the relic decaying particle parameter space that is interesting 
for the \li7, \li6 (and \be9) abundances.
The discussion will then prove useful in the following sections
for the identification of scenarios
in our particular model, supersymmetry with gravitino LSPs.
Fig.\ref{fig:banana1} shows the parameter space in relic decaying particle 
abundance, hadronic branching ratio $B_h$, and $X$-particle decay time which 
has impact on the cosmic lithium abundances while respecting all other 
abundance constraints. Note that results are only dependent on the product of 
$\Omega_Xh^2B_h$, where $\Omega_X$ is the fractional contribution of the
$X$-particle to the present critical density, if it wouldn't have decayed.
(The $\Omega_X$ may be converted to the X-particle number to 
entropy density $n_X/s$ via $n_XM_X/s = 3.6639\times 10^{-9}{\rm GeV}\Omega_X h^2$, where $M_X$ is the X-particle mass.)
Evidently the results depend as well on the hadronic branching ratio
of the X-particle decays. To produce the plots in Figs.\ref{fig:banana1} and
\ref{fig:banana2} we made the assumption of 
hadronic decays described by a $M_X = 1\, \rm TeV$ X-particle
decaying into a $q \bar{q}$ pair 
occuring with probability $B_h$.
This does not correspond exactly to the situation of 
NLSPs decays into 
gravitinos $\tilde{G}$, i.e. neutralino $\chi\to \tilde{G}q\bar{q}$
and stau $\tilde{\tau}\to \tilde{G}\tau q\bar{q}$, such that, for
our analysis, the figures
should be used only as guides.
However, making different assumptions for particle mass and initial
(hadronic) post-decay state make in many cases only changes 
of factors of order $2-3$. 
  
It is seen that for early decay times 
within a "banana"-shaped region the \li7 abundance may be significantly
reduced (\li7/H $< 2.5\times 10^{-10}$ red area). Here the upper envelope
of the area is defined by the constraint D/H$ \simle 5.3\times 10^{-5}$ and the
lower envelope by \li7/H $ \simge 2.5\times 10^{-10}$. A value of 
D/H $=5.3\times 10^{-5}$ may be already uncomfortably large. Therefore,
the accordingly labeled solid line 
in Fig.\ref{fig:banana1} indicates the less conservative 
D/H$< 4.0\times 10^{-5}$ limit. 
The green area indicates the region where a \li6/\li7 
ratio of $0.015\simle$\li6/\li7$\simle 0.3$ has been synthesized. 
Here the upper end of the range requires already some post BBN stellar \li6 
depletion (relative to \li7). The figure illustrates that 
for $\tau_X\simge 10^3$sec there exists
plenty of parameter space which may produce an observationally important
\li6 abundance by hadronic particle decays. Moreover,
as advocated in Ref.~\cite{Jedamzik:2004er}, a region
around $\tau_X\approx 1000\,$sec and $\Omega_Xh^2B_h\approx 2\times 10^{-4}$
may resolve both lithium anomalies at once.

In Fig.~\ref{fig:banana1} 
it has been implicitly assumed that the decaying particle is neutral.
In case it is charged, catalytic effects may have a strong impact on
the \li6 abundance~\cite{Pospelov:2006sc}, particularly for small $B_h$.
Moreover, catalytic effects may also lead to the production 
of \be9~\cite{Pospelov:2007js},
though it is currently not clear if this indeed happens 
(cf.~\cite{Kamimura:2008fx}).
Note that \be9 production is impossible with hadronic effects only. 
Fig.\ref{fig:banana2} shows the \li7, \li6, and 
(possibly) \be9 friendly areas for a charged 
massive particle decaying during the BBN era.  
Here a hadronic branching ratio $B_h = 10^{-4}$ and particle mass 
$M_X=1\,$TeV have been assumed. 
In order to compute \be9/H ratios we follow the assumptions 
in Ref.~\cite{Pospelov:2007js},
keeping in mind the possibility of very significant 
modifications~\cite{Kamimura:2008fx}.
It is noted that in Fig.\ref{fig:banana1} an ordinate
$\Omega_Xh^2B_h$ has been chosen, whereas in Fig.\ref{fig:banana2} the ordinate 
$\Omega_Xh^2$ is shown for one particular choice of $B_h$~\cite{remark1}. 
It is intruiging to remark that when the decaying particle is charged, not
only may the \li7 and \li6 anomalies be solved, but for the same parameters 
an observationally important 
$2\times 10^{-14}\simle$\be9$\simle 2\times 10^{-13}$ abundance 
\cite{Primas:2000gc, *Boesgaard:2005jf, *Boesgaard:2005pf}
may be synthesized as well. This is seen by the dashed lines
in Fig. \ref{fig:banana2} passing through the doubly preferred region
at $\Omega_Xh^2B_h\approx 3\times 10^{-4}$ and $\tau_X\approx 1500\,$sec.

Though the results of this section are essentially generic, and may be
used as guide lines for constructing other particular "lithium-friendly"
decaying particle scenarios, Fig.\ref{fig:banana1} also shows predictions within two particular 
setups in supersymmetric extensions of the standard model. The sequence of
grey points gives the prediction of stau-NLSPs in the CMSSM under the 
assumption of a gravitino mass of $m_{\tilde{G}} = 50\,$GeV. 
Such scenarios come very close to 
the doubly-preferred blue region at $\tau_X\approx  1000\,$sec, 
which illustrates that (within the approximations) the  \li6 and \li7 observed
abundances could be modified in an observationally favored way
by stau decay, albeit for a stau mass $m_{\tilde{\tau}}\sim 1\,$TeV 
unaccessible to the LHC.
This scenario will be analyzed 
once more in Section IV.
It is noted here that though it seems somewhat inconsistent to show
$\tilde{\tau}$ NLSP decay in Fig.~\ref{fig:banana1}, due to catalytic effects
which are not taken into account in that figure, conclusions concerning the
\li7 and \li7 {\it and} \li6 preferred regions are hardly modified when
catalytic effects are included (aside from \be9).
The sequence of blue points gives the prediction of neutralino NLSPs in
a GMSB model assuming $m_{\tilde{G}} = 100\,$MeV.  
Such NLSPs, characterized by fairly large freeze-out densities and
hadronic branching ratios, though they would produce too much \li6 when 
decaying late, could solve the \li7 problem for decay times 
$\tau_X\sim 100\,-400\,$sec. This possibility is studied in 
detail in Section V.

\section{Lithium and heavy LSP gravitinos}

\begin{figure}
\includegraphics[width=.45\textwidth]{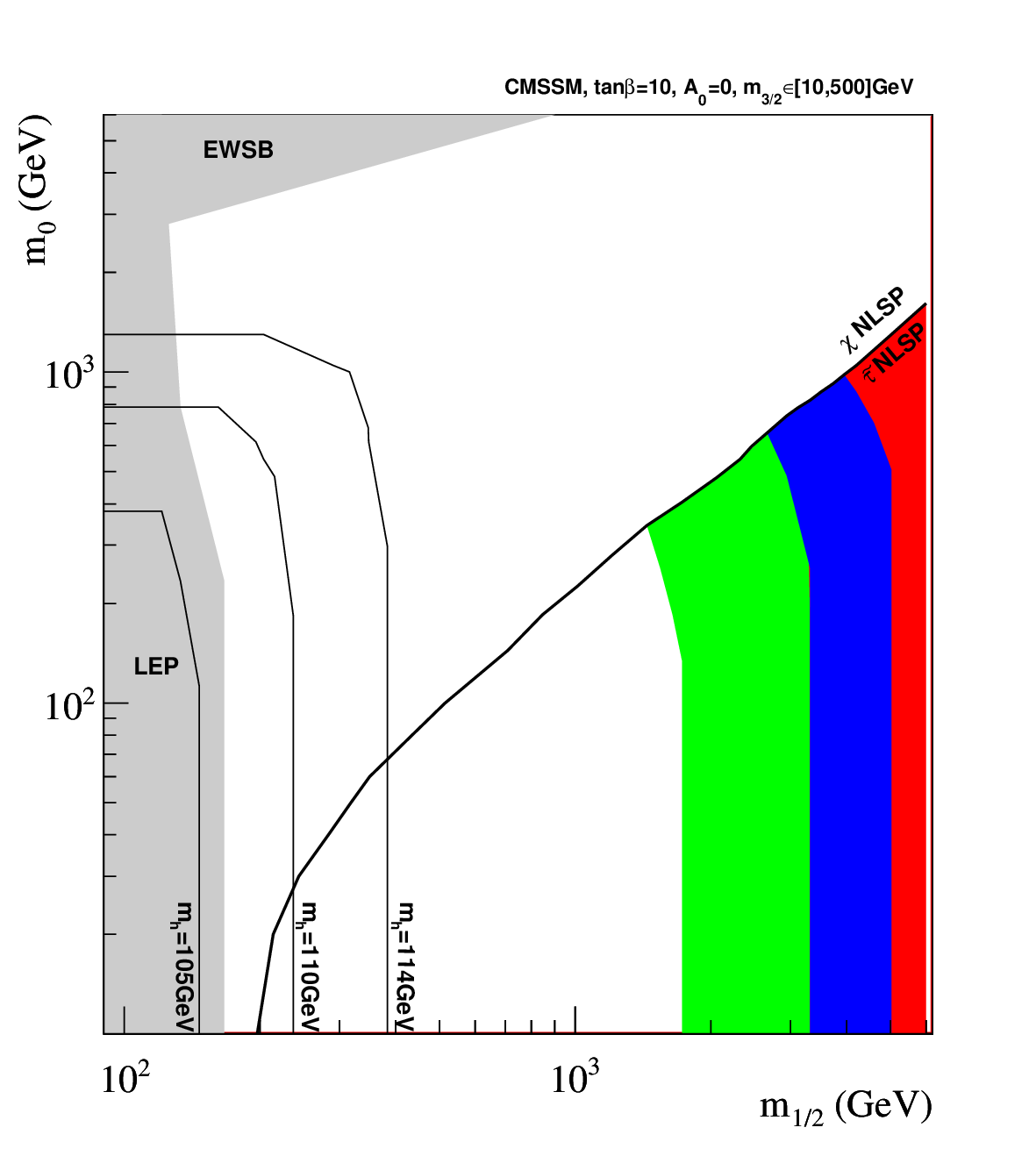}
\caption{Parameter space in the CMSSM 
$m_0 - m_{1/2}$ plane where NLSP decay during BBN results
in observationally favored modifications of the
\li7 and/or \li6 abundances with respect to standard BBN.
Here tan$\beta = 10$ has been assumed.
The color coding indicate: (red) \li7/H$< 2.5\times 10^{-10}$,
(green) $0.015<$\li6/\li7$<0.15$, and (blue) \li7/H$< 2.5\times 10^{-10}$
{\it and} $0.015<$\li6/\li7$<0.66$. Other constraints are as in 
Fig.~\ref{fig:banana1}. White areas are ruled out by BBN. 
The area labeled "LEP" is excluded by LEP accelerator constraints,
whereas the area labeled "EWSB" does not lead to electroweak symmetry
breaking. The line labeled "$\chi$ NLSP" and "$\tilde{\tau}$ NLSP" 
delineates the parameter space where a neutralino $\chi$ is the NLSP from
that where the stau $\tilde{\tau}$ is the NLSP. The three lines on the
left hand side give contours for the lightest neutral Higgs mass, as labeled.
}
\label{fig:m0m12_heavy}
\end{figure}

\begin{figure}
\includegraphics[width=.45\textwidth]{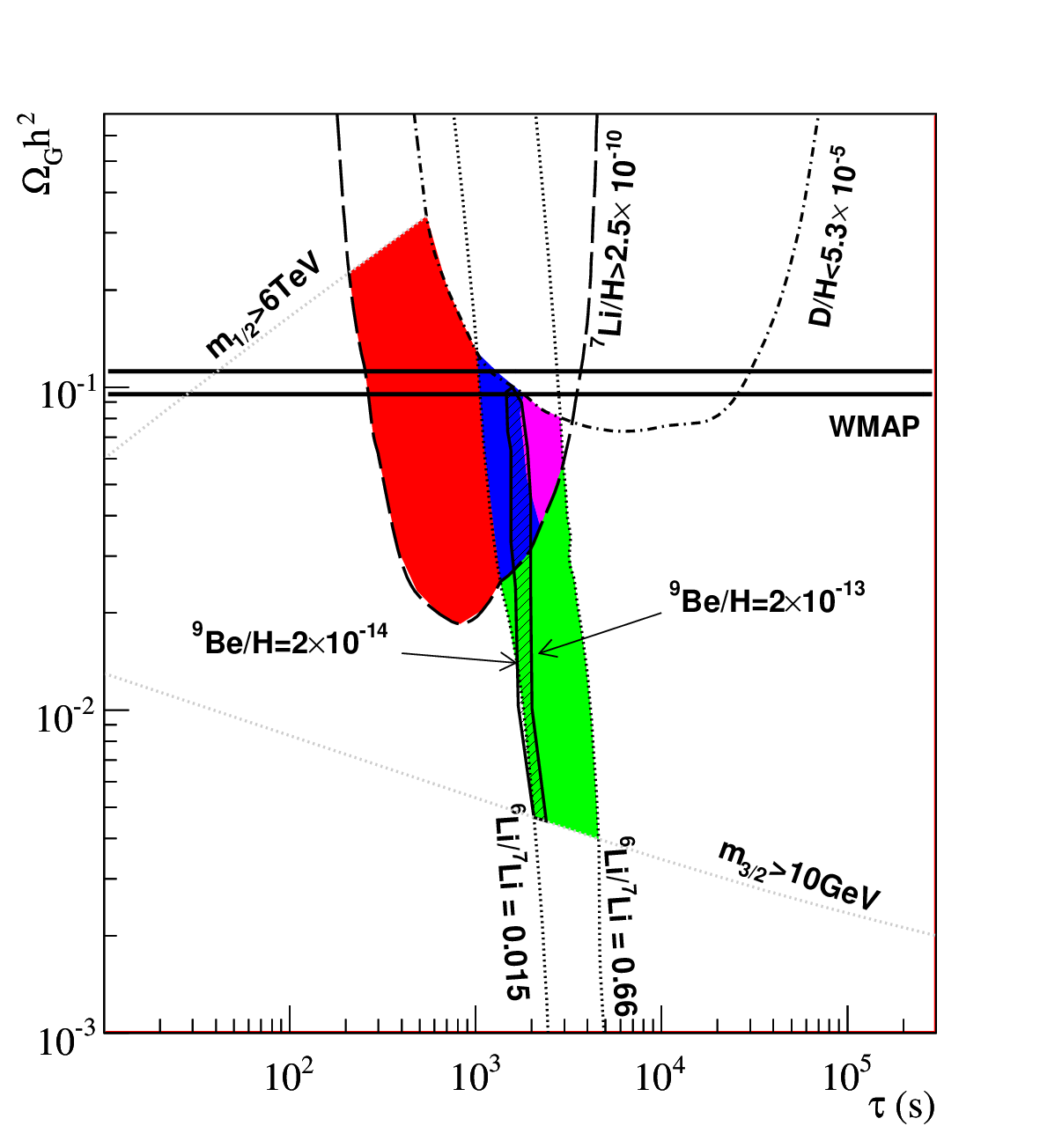}
\caption{Gravitino relic density $\Omega_{\tilde{G}} h^2$  
versus NLSP (stau) life time $\tau$ corresponding to the CMSSM models studied
in Fig.~\ref{fig:m0m12_heavy}. 
Color coding is as in Fig.~\ref{fig:m0m12_heavy} except that "pink" shows
regions where \li7/H$< 2.5\times 10^{-10}$ {\it and} 
$0.15<$\li6/\li7$< 0.66$ accounting for the possibility of \li6
depletion. The cross-hatched region shows parameter space where
$2\times 10^{-14}<$\be9/H$<2\times 10^{-13}$, though the estimate 
is uncertain. Other lines show (as labeled)
light element abundance constraints,
the scan range in $m_{1/2}$ and
$m_{\tilde{G}}$, as well as the by WMAP inferred range for the dark matter 
density.}
\label{fig:Omegatau_heavy}
\end{figure}

Weak scale mass \lq\lq heavy\rq\rq\, gravitinos are 
generally expected in the constrained minimal 
supersymmetric model (CMSSM), which will be utilised for the
present analysis with the gravitino mass, $m_{\tilde{G}} > 10\,$GeV
taken as a free parameter. Other parameters in the CMSSM are:
$m_0$, $m_{1/2}$ -- at the GUT-scale unified scalar, fermionic (gaugino) 
soft
supersymmetry breaking masses, respectively, $A_0$ -- trilinear couplings
in the scalar sector (assumed to be zero throughout), and 
tan$\beta$ -- the ratio of the vacuum expectation values of the two Higgs
required in supersymmetry.
It is known 
that the CMSSM with weak-scale LSP gravitinos
may lead to cosmologically interesting changes in the primordial
\li6 and \li7 abundances when the superpartner to the tau lepton, 
the stau, is the NLSP~\cite{Jedamzik:2004er,Jedamzik:2005dh,Cyburt:2002uv}.
This may be seen given their abundances, which can be approximated for
moderate $\tan \beta$ by  \cite{Asaka:2000zh}
\begin{equation}
\Omega_{\tilde{\tau}}h^2 \approx (2.2 - 4.4) \times 10^{-1}
\left(\frac{m_{\tilde{\tau}_1}}{1 \rm TeV}\right)^2\, ,  
\label{eq:staurelic}
\end{equation}
their typical hadronic branching ratios
$B_h\approx 10^{-4} - 10^{-2}$,
and life times in the range 
$\tau_{\tilde{\tau}} \approx 10^2 - 10^5 \rm sec$ as controlled by
Eq.(\ref{eq:lifetime}) with $\kappa =1, n=4$. Note that the range in 
the approximation Eq.(\ref{eq:staurelic}) takes into account possible 
co-annihilation with the other sleptons  (but does not account for
co-annihilation with a neutralino or other specific effects at
larger $\tan \beta$, such as mixings, higgs resonances, enhanced
couplings, etc.). 
When compared to Fig.~\ref{fig:banana1} and taking $\tan \beta =10$, 
one finds for instance that 
staus in the mass range $m_{\tilde{\tau}}\approx 1-1.5\,$TeV, 
having $B_h \simge 10^{-3}$
and decaying into gravitinos with 
$m_{\tilde{G}}\approx 50-200\,$GeV pass 
right through the preferred (blue) region. A somewhat heavier stau
$\approx 2\,$TeV, with a $B_h \approx 2\times 10^{-3}$ and a
$\Omega_{\tilde{\tau}}h^2 B_h \approx 1\times 10^{-3}$, would be {\it only} 
\li7 friendly as can be seen from Fig.~\ref{fig:banana1}, for
$m_{\tilde{G}}\approx 100\,$GeV.
The lithium friendly
parameter space for the CMSSM with $\tan\beta = 10$
is shown in Figs.~\ref{fig:m0m12_heavy} and
~\ref{fig:Omegatau_heavy}. 
The calculations leading to 
Figs.~\ref{fig:m0m12_heavy} and~\ref{fig:Omegatau_heavy} include the 
improvements (1)-(3),(5), and (6)
of the list given in the Introduction, 
but not improvement (4). Treating each case
with a realistic nucleon spectrum is numerically too expansive, and will be
only done for a few models below. 
Note that Figs.~\ref{fig:m0m12_heavy} and~\ref{fig:Omegatau_heavy}
really show results of higher-than-two dimensional parameter space.
Even at fixed $\tan\beta$ and $A_0$ the results in the $m_0-m_{1/2}$ plane
shown in Fig.~\ref{fig:m0m12_heavy} are for a variety of gravitino masses. 
In particular, blue areas in the $m_0-m_{1/2}$ plane are covering up green and
red areas which result for different choices of $m_{\tilde{G}}$.
The figure, and other
figures which follow, thus show where \li6 (green), \li7 (red), and
\li6 and \li7 (blue) friendly regions are expected, 
{\it in case} $m_{\tilde{G}}$ has
been approximately chosen. In contrast, in white regions 
for no $m_{\tilde{G}}$ within the adopted range may light element abundance
constraints be met.

Figs.~\ref{fig:m0m12_heavy} and~\ref{fig:Omegatau_heavy} may be directly compared to Figs. 1 and 2
of \cite{Jedamzik:2005dh}. It is seen that the doubly preferred (blue) region is narrower
in \cite{Jedamzik:2005dh} compared to the present study. This is simply due to 
a finer sampling of the gravitino mass in the present paper. Furthermore, models with larger
stau life times, $\tau_{\tilde{\tau}}\simge 5\times 10^3$sec, are due to
catalytic \li6 and (possibly) 
\be9 overproduction now ruled out~\cite{Pradler:2007is,Pospelov:2008ta}. 
Finally, it is noted that in the
$\Omega_{3/2}h^2$ - $\tau_{\rm NLSP}$ plane the doubly preferred (blue) region
has hardly moved. This is somewhat surprising since for $\sim 2$
larger $B_h$ (point 2 in Introduction), as is the case, 
one would expect for the region to move
a factor $\sim 2$ lower in $\Omega_{\tilde{\tau}}h^2$
(and thus $\Omega_{3/2}h^2$) as to not overproduce D.
Nevertheless, this effect is counter-balanced by a lower effective 
$\langle m_{q\bar{q}}\rangle$ (point 3 in Introduction), 
implying less \lq\lq distortion\rq\rq of
the light elements than initially envisioned.

Scenarios where stau NLSPs decay at around $\tau\approx 1000\,$sec
into gravitinos, thereby solving both lithium problems at once, have the added, 
and totally accidental, benefit of coming tantalizingly close to 
producing all the dark matter in form of warm gravitinos during the decays. 
This is nicely illustrated in Fig.~\ref{fig:Omegatau_heavy} where the blue 
area just overlaps from below with the WMAP strip,  
and models at higher $\Omega_{3/2}h^2$ are ruled out by Deuterium
overproduction, D/H$> 5.3\times 10^{-5}$. 
It is therefore interesting to see if this conclusions survives
when a calculation of improved accuracy is performed. 
The results in Fig.~\ref{fig:Omegatau_heavy} 
rely on  approximating the "hadronic energy" 
by $\langle m_{q\bar{q}}\rangle$ of 
Eq.(\ref{eq:SteffenEHad}). 
When properly calculating the hadronic energy release 
employing a realistic energy distribution of the nucleons 
leads to a milder D overproduction constraint
than expected from the $\langle m_{q\bar{q}}\rangle$ approximation
as shown in section II, thus pushing the blue area more
upwards into the WMAP strip.
This is seen in Fig.~\ref{fig:points}
where results for CMSSM ($\tan\beta = 10$) scenarios with 
$\Omega_{\tilde{G}}h^2$ close to $0.1$ are shown, which have been computed
without the $m_{q\bar{q}}$ approximation.
Such scenarios are thus {\it fully consistent} with producing
{\it all} the dark matter non-thermally.
 
It is of interest if catalytic effects due to the electrically charged 
staus may also
lead to a cosmologically important \be9 abundance \cite{Pospelov:2007js,
Pospelov:2008ta} in the doubly
preferred parameter space. Indeed, this is the case, as may be seen
in  Fig.~\ref{fig:Omegatau_heavy} where the \be9/H -ratio in the
range $2\times 10^{-14} \simle \mbox{\be9/H} \simle 2\times 10^{-13}$
is indicated by the cross-hatched region. 
This is indeed very interesting since the observed \be9/H in the lowest
metallicity stars is 
\be9/H$\approx 
3\times 10^{-14}-10^{-13}$~\cite{Primas:2000gc, *Boesgaard:2005jf, 
*Boesgaard:2005pf}, thus close to the predicted one. 

In Table I parameters and BBN yields for some particular SUSY points in
the CMSSM (and GMSB, cf. Section V) for $\tan\beta = 10$ are shown. 
One notes two classes of models, those which synthesize much \li6 (and \be9),
for larger $\tau_X$, and those which significantly reduce \li7 for 
smaller $\tau_X$, with the latter models also accounting for the dark matter
produced during the decay.
In either model D/H may be significantly less than 
$4\times 10^{-5}$. Models which accomplish both, a factor $> 2$ destruction 
of \li7 as well as cosmologically interesting production of \li6 
(and possibly \be9), however, typically tend to 
have D/H$> 4\times 10^{-5}$.
Finally, it is noted that even in the presence
of hadronic decays, i.e. injection of extra neutrons, the \be9/\li6 ratio
falls typically in the range $1-4\times 10^{-3}$, as proposed 
in Ref.~\cite{Pospelov:2007js} in the absence of extra neutrons.

Last but not least Fig.~\ref{fig:m0m12_heavy_tb50} shows results for the CMSSM and 
$\tan\beta = 50$. As was already noted in \cite{Jedamzik:2005dh}, for 
$\tan\beta$ large, and a stau NLSP (for neutralino NLSPs cf. Section V), typical 
$\Omega_{\tilde{\tau}}h^2B_h(\tilde{\tau})\simle 10^{-4}$ are too small to
resolve the \li7 problem (cf. Fig.~\ref{fig:banana1}). This is due to the
smaller freeze-out abundance due to efficient annihilations of 
$\tilde{\tau} \bar{\tilde{\tau}}$ into $W^{+} W^{-} (ZZ)$ and $hh$ (light CP-
even Higgses) for large $\tan \beta$, as compared to the $\tilde{\tau} \tilde{\tau}
\to \tau \tau$ channel which controls the abundance at low $\tan \beta$.
Such effects can be due to Higgs poles or large left-right stau mixing 
 for relatively light stau masses 
 (as shown in \cite{Ratz:2008qh, Pradler:2008qc} in non-CMSSM scenarios).
 In the CMSSM and for very heavy stau NLSPs as is the case in  
 Fig.~\ref{fig:m0m12_heavy_tb50},
 $W^{+} W^{-}$  plus $ZZ$ channels become dominant. 
Nevertheless, there remains parameter space 
which may synthesize observationally important primordial \li6 
and \be9 abundances at once. This may occur for staus as light as 
$m_{\tilde{\tau}} \approx 500 {\rm GeV}$, potentially visible at the LHC.
It is noteworthy that for very heavy $m_{\tilde{\tau}} \simge 2 {\rm TeV}$
there are again regions where either \li7 alone or both \li6 and \li7 are
synthesized at the observationally inferred level. Furthermore, 
due to the smallness of $\Omega_{\tilde{\tau}}h^2$ for relatively light
staus, only a small fraction of the dark matter would be produced by 
stau decays, with the "missing" gravitinos possibly produced during reheating at
a comfortably large reheat temperature of $T_{RH}\approx 10^{10}$GeV.
Very heavy stau decays can still account for all the (gravitino) dark matter
as shown in  Fig.~\ref{fig:m0m12_heavy_tb50},
nevertheless models at smaller $\tan\beta \sim 10$ seem more economical in the 
context of dark matter generation.

\bef
\includegraphics[width=.45\textwidth]{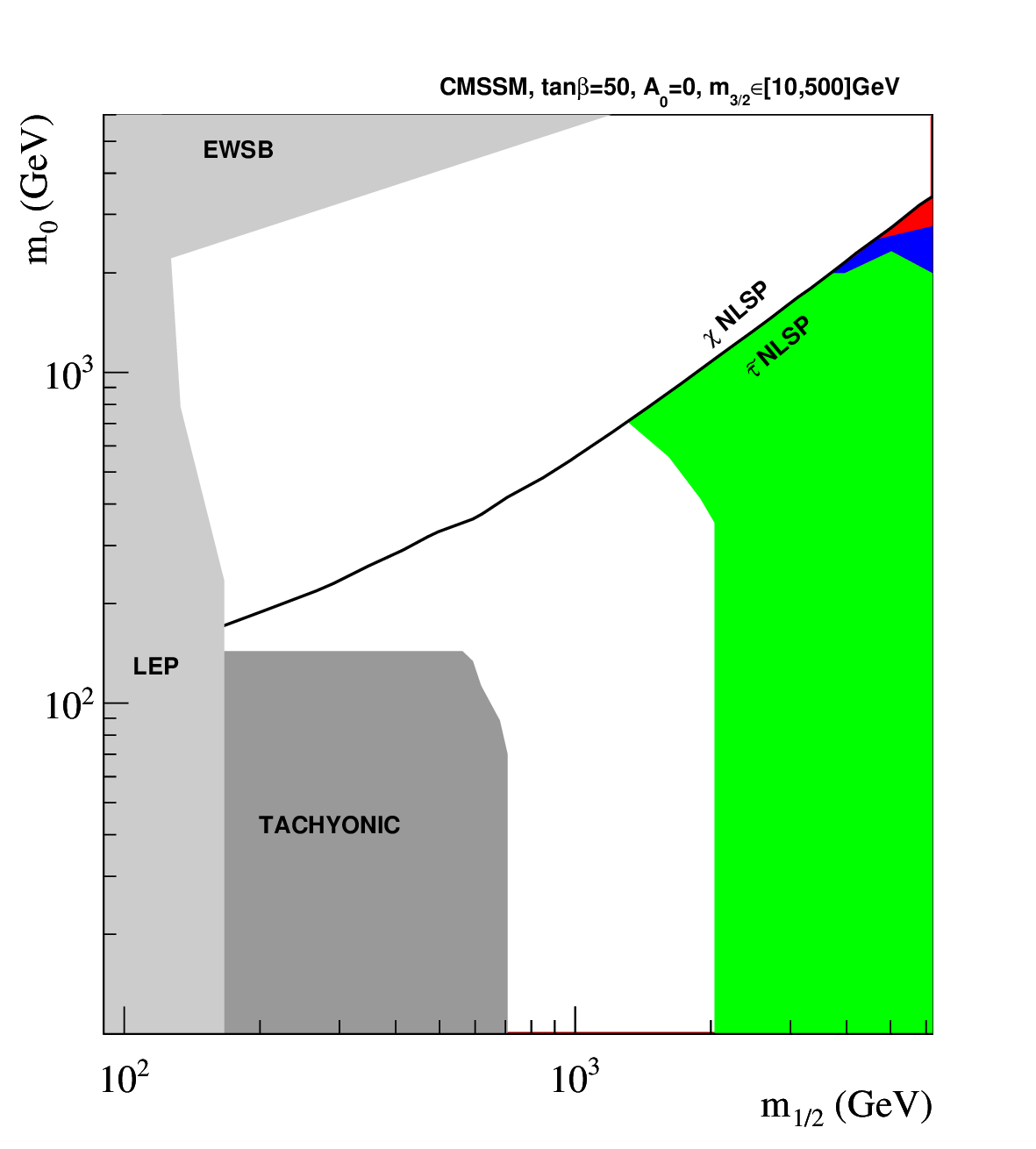}
\includegraphics[width=.45\textwidth]{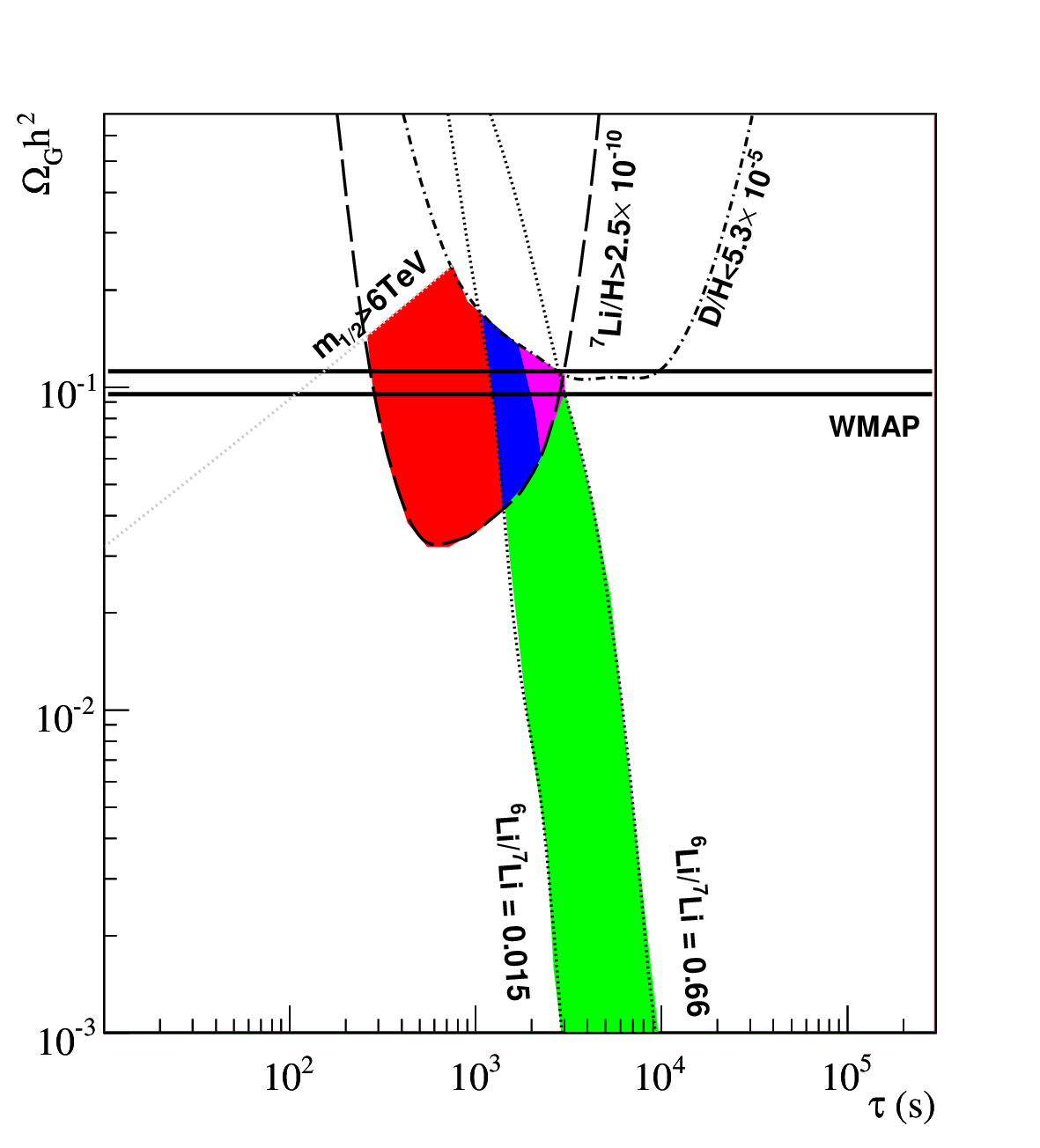}
\caption{Same as Figs.\ref{fig:m0m12_heavy} and \ref{fig:Omegatau_heavy}, 
but for $\tan \beta = 50$. The region labeled "tachyonic" is excluded
due to some sfermion masses becoming negative.}
\label{fig:m0m12_heavy_tb50}
\eef

\begin{footnotesize}
\begin{table}
\newcommand{\lstrut}{{$\strut\atop\strut$}}
\caption{Light-element abundances yields and gravitino abundance
$\Omega_{3/2}h^2$ for a number of selected models. Cf. to Table II
for the particular particle model parameters. Model denoted by $(a)$ have
been computed with the invariant mass approximation (see text for details).}
\label{Table:I}
\vspace{2mm}
\begin{center}
\begin{tabular}{|c||c|c|c|c|c|}
\hline
Model & D/H & \li7/H & \li6/\li7 & \be9/H & $\Omega_{3/2}h^2$ \\
\hline

A & $3.45\cdot 10^{-5}$ & $2.20\cdot 10^{-10}$ & -- & -- & $0.11$ \\

A$^{a}$ & $3.86\cdot 10^{-5}$ & $1.66\cdot 10^{-10}$ & -- & -- & $0.11$ \\

B & $4.75\cdot 10^{-5}$ & $2.14\cdot 10^{-10}$ & 0.058 & $3.4\cdot 10^{-14}$ &
$0.10$ \\

B$^{a}$ & $5.42\cdot 10^{-5}$ & $1.83\cdot 10^{-10}$ & 0.081 & $3.8\cdot 10^{-14}$ &
$0.10$ \\

C$^{a}$ & $3.59\cdot 10^{-5}$ & $3.32\cdot 10^{-10}$ & 0.044 & $7.5\cdot 10^{-14}$ &
$0.0086$ \\

D$^{a}$ & $2.57\cdot 10^{-5}$ & $4.95\cdot 10^{-10}$ & 0.044 & $1.9\cdot 10^{-13}$ &
$3\cdot 10^{-4}$ \\

E$^{a}$ & $3.61\cdot 10^{-5}$ & $2.09\cdot 10^{-10}$ & -- & -- &
$2\cdot 10^{-4}$ \\

\hline
\end{tabular}
\end{center}
\end{table}
\end{footnotesize}

\begin{footnotesize}
\begin{table}
\newcommand{\lstrut}{{$\strut\atop\strut$}}
\caption{Particle physics model parameters corresponding to the
abundance yields shown  in Table I. All masses are in GeV and 
$\tan\beta=10$}
\label{Table:II}
\vspace{2mm}
\begin{center}
\begin{tabular}{|c||c|c|c|c|c|}
\hline
Model & SUSY - NLSP & $M_{NLSP}$ & $m_{\tilde{G}}$ & $\tau_X$ (s) \\
\hline

A & CMSSM $\tilde{\tau}$ & $2060$ & $160$ & $403$ \\
  & $(m_0,m_{1/2})=(1045,5020)$ & & & \\

B & CMSSM  $\tilde{\tau}$ & $1638$ & $175$ & $1553$ \\
  & $(m_0,m_{1/2})=(678,4200)$ & & & \\

C & CMSSM  $\tilde{\tau}$ & $763$ & $30$ & $1980$  \\
  & $(m_0,m_{1/2})=(395,1831)$ & & & \\

D & GMSB $N=2$ $\tilde{\tau}$ & $264$ & $2.72$ & $3320$  \\
  & $(\Lambda,M_{\rm mess})=(10^5,5\cdot 10^6)$ & & & \\

E & GMSB $N=1$ $\chi$ & $133.8$ & $0.08$ & $103$  \\
  & $(\Lambda,M_{\rm mess})=(10^5,5\cdot 10^6)$ & & & \\

\hline
\end{tabular}
\end{center}
\end{table}
\end{footnotesize}

\section{Lithium and light gravitinos}

In this section we identify SUSY parameter space which
results in significant changes of the primordial \li6 and \li7 abundances,
in cases when the gravitino is rather light $m_{\tilde{G}}\simle 10\,$GeV.
Such an analysis has so far not been performed. Light gravitinos are
a typical prediction
of gauge-mediated SUSY breaking scenarios (GMSB), where SUSY
breaking in a hidden sector is transmitted to the visible sector via
gauge interactions of some messenger fields \cite{Fayet:1978qc,
*Dine:1981za, *Dimopoulos:1981au,*Dine:1981gu, *Dine:1982qj, *Dine:1982zb,
*AlvarezGaume:1981wy, *Nappi:1982hm, *Dimopoulos:1982gm}, 
\cite{Dine:1993yw, *Dine:1994vc,*Dine:1995ag}. Assuming a 
typical grand unified group,
such models have three continuous and one discrete parameter, 
namely $\Lambda$,  the 
common mass scale of the soft Susy breaking masses, $M_{\rm mess}$, the Susy 
preserving messenger mass,  $\tan \beta$, the ratio of the two
higgs doublet vevs and $N_{mess}$  
the number of quark-like (assumed to be equal the number of
lepton-like) multiplets
(of some GUT group) messenger supermultiplets.
The soft Susy breaking gaugino (square of scalar) masses are then 
generated at the $1$- ($2$-) loop level, respectively,
leading to the physical spectra and
couplings of the MSSM particles.
 (The soft breaking trilinear
couplings $A$ are generated only at the two-loop level and will be
set to zero).  Even though the fine details of
the MSSM spectrum and couplings depend on these four parameters, in most
parts of the parameter space varying $M_{\rm mess}$, which affects 
the soft masses only logarithmically, 
will be effectively irrelevant to our study. 
We are thus left with one free mass parameter $\Lambda$ which should
furthermore be of order $100\,$TeV if the MSSM spectrum is to remain
at the electroweak mass scale. 
Such models are therefore more constrained than, for
example, the CMSSM which has (at least) two mass parameters $m_0$ and $m_{1/2}$
(given that trilinear couplings $A$'s have been set to zero). 
Nevertheless, before discussing results in the GMSB, we will still study some
aspects of the CMSSM when the gravitino mass 
is (arbritrarily) low. Though not quite consistent, as one
expects $m_{\tilde{G}}\sim m_{soft}$ when SUSY breaking in a hidden
sector is communicated to the visible sector by gravitational 
interactions, the study of the CMSSM with light gravitinos 
may help to shed
more light on existing 
"lithium-friendly" parameter space in models 
beyond the toy models CMSSM and GMSB.

Consulting Fig.~\ref{fig:banana1} as the key figure of where to expect 
parameter space solving the \li7 anomaly, one observes that for all
$3\times 10^{-4}\simle \Omega_{\rm NLSP}h^2B_h\simle 0.1$, and with a 
somewhat tuned gravitino mass to match the "desired" NLSP life time,
solutions should exist. It is noted here that the exact range required
in $\Omega_{\rm NLSP}h^2B_h$ depends on the NLSP mass
$M_{\rm NLSP}$, moving down as
$M_{\rm NLSP}$ moves down (Fig.~\ref{fig:banana1} assumes 
$M_{\rm NLSP} = 1\,$TeV) and vice versa.
As discussed above, stau NLSP abundances 
are not much larger than $\Omega_{\tilde{\tau}}h^2\sim 1$ 
(cf. Eq.~\ref{eq:staurelic}), 
whereas their hadronic branching ratios are usually small $B_h\simle 10^{-3}$. 
Thus, \li7 solving areas for the stau maybe only found for large 
$m_{\tilde{\tau}}\sim 1\,$TeV since only there 
$\Omega_{\tilde{\tau}}h^2\sim 1$. Furthermore, to match the desired life
time, large $m_{\tilde{\tau}}$ require large $m_{\tilde{G}}$ (cf. Eq.~(\ref{eq:lifetime})), 
thus pointing towards gravity mediated susy breaking scenarios.
These solutions have been all discussed
in Section IV. In contrast, the freeze-out $\Omega_{\chi}h^2$ for
neutralinos may be large $\sim 1-100$ and there hadronic branching ratio
due to $\chi\to\tilde{G}q\bar{q}$ with an intermediate $Z$ or photon is
typically of the order of $B_h\sim 0.1$. In particular, one may find 
\li7-friendly models for comparatively light neutralinos only for which
$\Omega_{\chi}h^2$ and $B_h$ are not too large as to saturate the upper
bound
on $\Omega_{NLSP}B_h(NLSP)$ given by \he4 overproduction. 
Since $\tau_{\chi}$ scales with
$m_{\chi}$ to the fifth power, the gravitino
mass $m_{\tilde{G}}$ in such decays has to be rather light in order to match
the decay time window $100\, {\rm sec}\simle\tau\simle 1000\, {\rm sec}$
(cf. Fig.~\ref{fig:banana1}). Thus, one is automatically led to lighter gravitinos.
This implies also, since $\Omega_{NLSP}$ may not be too large, and
$\Omega_{3/2}=\Omega_{NLSP}(m_{\tilde{G}}/m_{NLSP})$ that only a small fraction of
the dark matter would be created by such decays.

\begin{figure}
\includegraphics[width=.45\textwidth]{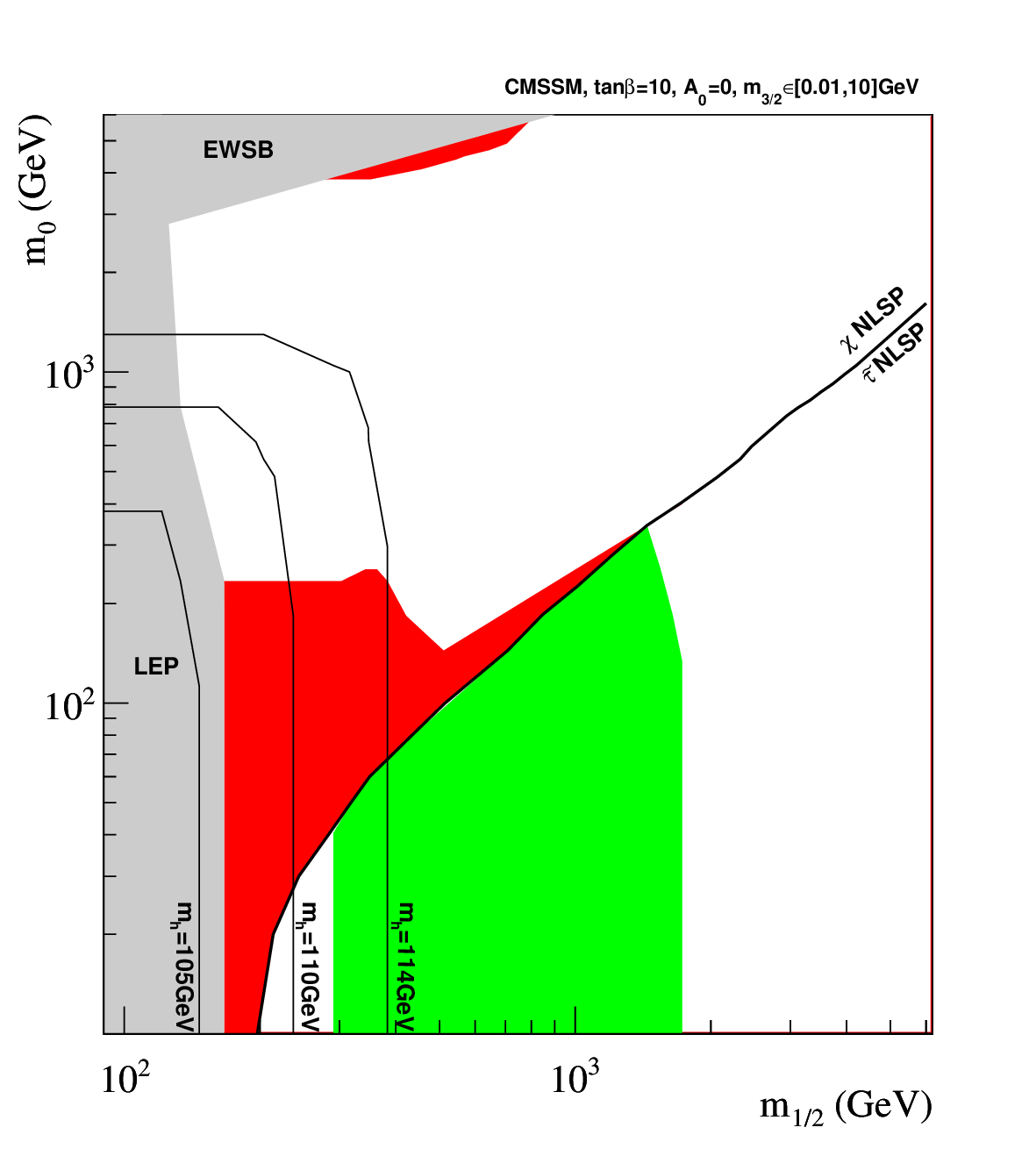}
\caption{\li7 (red) and \li6 (green) solving regions for a CMSSM model 
at $\tan\beta = 10$ and with light gravitinos in the range
$10\,{\rm MeV}\simle m_{\tilde{G}}\simle 10\, {\rm GeV}$. All constraints
and lines have similiar meaning to those in Fig.~\ref{fig:m0m12_heavy}. 
Note that
the \li7 solving regions exist essentially whereever 
$3\times 10^{-4}\simle \Omega_{\rm NLSP}h^2B_h({\rm NLSP})\simle 0.1$.}
\label{fig:m0m12_light}
\end{figure}

\bef
\epsfxsize=8.5cm
\includegraphics[width=.45\textwidth]{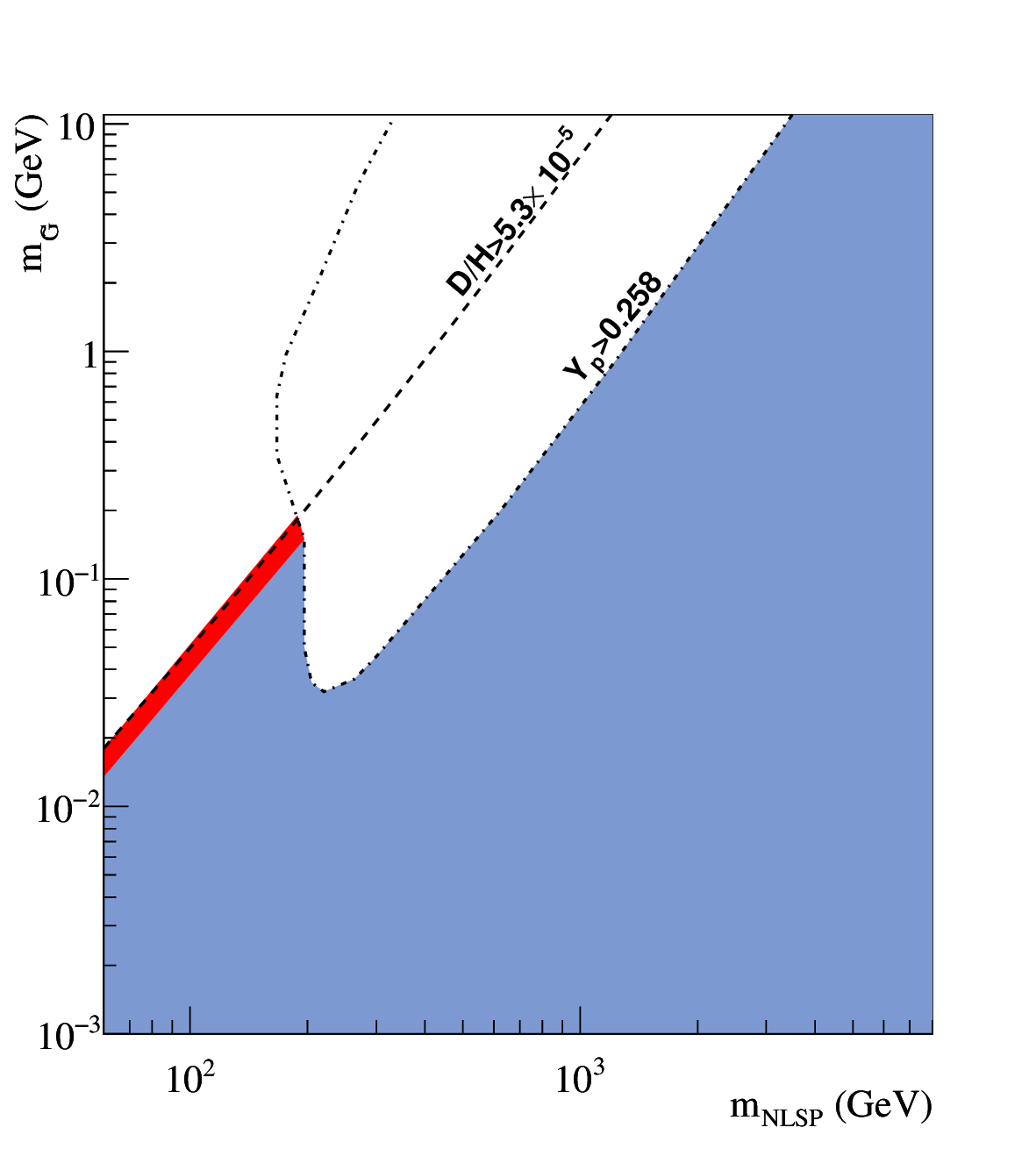}
\caption{Parameter space in the gravitino mass $m_{\tilde{G}}$
NLSP mass $m_{\rm NLSP}$ plane for light gravitinos $m_{\tilde{G}}<10\,$GeV
which significantly reduce the \li7 abundance, 
i.e. \li7/H$<2.5\times 10^{-10}$ (red), while respecting all other BBN
constraints, as labeled. Here a GMSB model with $\tan \beta = 10$,
$N_{mess}=1$, and $M_{mess}= 5 \times 10^6 {\rm GeV}$ has been assumed. 
Light-blue areas are allowed by BBN while white areas are ruled out.}
\label{fig:GMSB1}
\eef

\bef
\epsfxsize=8.5cm
\includegraphics[width=.45\textwidth]{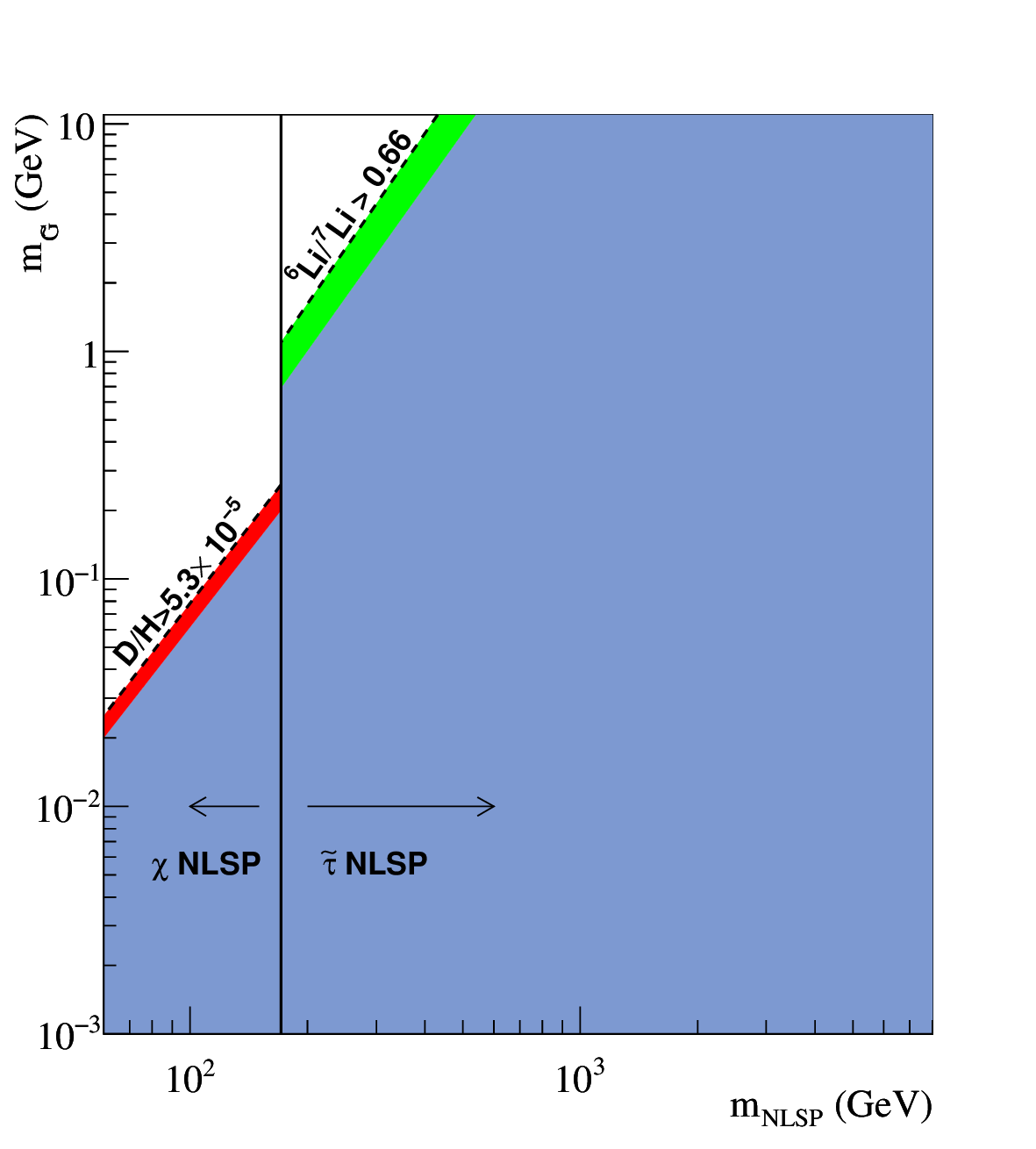}
\caption{As Fig.~\ref{fig:GMSB1} but with $N_{mess}=2$, leading to
neutralino NLSPs for $m_{\rm NLSP}\simle 175\,$GeV and stau NLSPs for
$m_{\rm NLSP}\simge 175\,$GeV as indicated. The green area shows regions with
$0.015\simle$\li6/\li7$\simle 0.66$.}
\label{fig:GMSB2}
\eef

Identifying areas of small neutralino abundance 
in the CMSSM
in the $m_0$-$m_{1/2}$ plane, in order to satisfy
$\Omega_{NLSP}B_h(NLSP)\simle 0.1$ (rather 
$\Omega_{NLSP}B_h(NLSP)\simle 0.03$ for $m_{\chi}\approx 100\,$GeV),
is reminescent of identifying the regions where 
$\Omega_{\chi}\sim \Omega_{WMAP}$ to account for neutralino dark matter. 
The latter question has received much attention
over the last years. It is therefore not surprising that one finds
potential \li7-solving parameter space in three distinct 
and "known" regions
(a) the bulk region close to the LEP bound at small 
$m_{1/2}\approx 300-400\,$GeV, (b) the co-annihilation region close 
to the line of neutralino-stau mass degeneracy, and (c) the focus point
region at large $m_0$ and $m_{1/2}$. 
These regions are shown in red in Fig.~\ref{fig:m0m12_light} for 
$\tan \beta=10$. In contrast with the heavy gravitino case, here Higgs
mass bounds from LEP  could potentially exclude large 
fractions of the bulk region, as illustrated  in the figure with  
lines of constant lightest Higgs mass including the Standard Model
Higgs present lower limit $m_{H^0}\simge 114\,$GeV. 
Although the experimental limits from LEP and
Tevatron are milder for the MSSM higgses, one should 
keep in mind that a detailed study is needed, beyond the maximal mixing
or no-mixing assumptions, in order to assess quantitatively 
the effect of experimental exclusions on our scenarios.
There exists more parameter space to solve 
the \li7 problem by neutralino decay than to produce neutralino dark matter
of the right density (when $m_{\tilde{G}} > m_{\chi}$). 
Here region (a) requires gravitino masses 
$30 \rm MeV \simle m_{\tilde{G}}\simle  100 \rm MeV$, region 
(b) $ 200 \rm MeV \simle m_{\tilde{G}}\simle  6.5 \rm GeV $,
whereas region (c) prefers 
$ 200 \rm MeV \simle m_{\tilde{G}}\simle 800 \rm MeV$.
Note also that for larger $\tan \beta$  values ($\simge 40$)  a fourth \li7 friendly funnel 
shaped region appears, corresponding to the well-known CP-odd higgs resonance 
effects. 

We next study GMSB models. The blue sequence of points
in Fig.\ref{fig:banana1} shows the prediction for $\Omega_{\chi} h^2B_h$ for NLSP
neutralinos decaying into $100\,$MeV gravitinos in a 
particular GMSB model, employing $N_{mess}=1$ (lepton- and quark-like) 
messenger particles of mass
$M_{mess} = 5\times 10^6$ GeV  and $\tan\beta = 10$. 
More accurate results for this model, now
with varying $m_{\tilde{G}}$, are shown in Fig.~\ref{fig:GMSB1}. 
There clearly exists \li7-solving parameter space for light neutralinos 
$\simle 200$GeV detectable at the LHC. 
We have also shown in light blue the parameter space region consistent with 
standard BBN (whereas the white region is ruled out). 
This is interesting by itself, as it illustrates the overall
consistency of GMSB scenarios with primordial nucleosynthesis irrespective
of whether the solution to the \li7 problem is of particle physics or other
astrophysical origin. In particular a combination of the standard bounds
on \he4 mass fraction and Deuterium abundance cuts all  gravitino masses
 $\simge 200\,$MeV for a neutralino NLSP in the "natural" range $\simle 1$ TeV,
 thus reinforcing the common lore of light gravitinos within GMSB.   
Finally, we illustrate in Fig.~\ref{fig:GMSB2} a GMSB scenario with 
two messenger (super)multiplets, $N_{mess}=2$. 
In this case, the NLSP can be a stau or
a neutralino, depending on the values of $\Lambda$. Here a \li6-solving pattern
requires a stau NLSP heavier than $\sim 175\,$GeV with $m_{\tilde{G}} \simge 1$
GeV, while solutions for \li7 still obtain for a neutralino NSLP in the same
part of the parameter space as in the $N_{mess}=1$ case. The discontinuity in  
Fig.~\ref{fig:GMSB2} is obviously due to the different nature of the NLSP,
different hadronic branching ratios as well as absence of catalysis effects for
neutralino NLSP. We note also that an intermediate range of gravitino masses
is disfavoured if one insists on solving at least one of the two lithium
problems on top of the consistency with standard BBN. However this gap in the 
mass range is obtained for a fixed messenger mass. We checked that it can be 
easily filled by  varying the latter over a few orders of magnitude. Other
patterns can appear by varying the number of messengers, or by relaxing the
grand unification assumptions taking different numbers of quark-like and
lepton-like messenger fields.

\section{Conclusions}

\begin{footnotesize}
\begin{table}
\newcommand{\lstrut}{{$\strut\atop\strut$}}
\caption{Potential for SUSY with gravitino LSPs to resolve the
\li7 problem, account for \li6, produce \be9, account completely 
for the dark matter due to non-thermal decay production, and be 
detectable at the LHC.}
\label{Table:III}
\vspace{2mm}
\begin{center}
\begin{tabular}{|c|c|c|c|c|c|c|}
\hline
Gravitino & NLSP & \li7 & \li6 & \be9 & $\Omega_{\rm DM}h^2$ &
LHC \\
\hline
light & stau & X & $\surd$ & $\surd$ ? & X & $\surd$ \\
      & neutralino & $\surd$ & X & X & X & $\surd$ \\
\hline
heavy & stau & $\surd$ & $\surd$ & $\surd$ ? & $\surd$ & X \\
      & neutralino & X & X & X & X & $\surd$ \\
\hline
\end{tabular}
\end{center}
\end{table}
\end{footnotesize}

The present paper has studied a variety of supersymmetric scenarios
with gravitino LSPs. It has been mostly motivated by an apparent mismatch 
between the by standard BBN (SBBN) predicted \li7/H ratio and that observed in 
low-metallicity halo stars, as well as the (potentially controversial)
claims of an observed \li6/\li7 plateau at low metallicity reminiscent
of a \li6 primordial abundance factor $\sim 10^3$ larger than predicted
by SBBN. Though one or both of these anomalies may have 
an astrophysical/observational
origin, it is conceivable that they point to physics beyond the 
standard model of particle physics and BBN.
In particular, supersymmetric (NLSP) particle decays during BBN may 
explain either one or both of these anomalies.
Though not the first such study, the present study improves
in accuracy on several accounts, as well as studies, for the first time, 
the case of light gravitinos 
$10\, {\rm MeV}\simle \m_{\tilde{G}}\simle 10 {\rm GeV}$. 
Detailed and improved analyses of relic NLSP abundances
and hadronic branching ratios are performed in two distinct SUSY models,
the CMSSM with heavy gravitinos $m_{\tilde{G}}\simge 10\,$GeV 
(Section IV) and
the GMSB with light gravitinos $m_{\tilde{G}}\simle 10\,$GeV (Section V). 
Other significant improvements concern the study of BBN yields due to
NLSP hadronic decays when no simplifying assumptions are made about
the injected nucleon spectrum (Section II), as well as the full inclusion
of catalytic effects during BBN due to the presence of weak-mass scale
electrically charged particles (i.e. the stau). Finally, approximate
"model builders instructions" of where, on general grounds, 
to find relic particle decay parameter space relevant to 
the \li7 and \li6 abundances are given as well (Section III).   

The general results of this study are summarized in Table III.
Susy with either heavy or light gravitino LSPs may
impact the cosmic lithium abundances while satisfying all other BBN
constraints. Scenarios with heavy gravitinos require the charged stau
to be the NLSP, and may solve the \li7 problem with or without production 
of \li6 in abundance as claimed to be observed in low-metallicity
stars, while producing all the dark matter as inferred by WMAP.
These conclusions are unchanged from earlier ones, independent of
catalysis effects and improvements in hadronic branching ratios and
injected nucleon spectrum due to the decay. However, in those
parts of the parameter space where significant \li6 is synthesized,
interestingly, it is conceivable 
that \be9 on levels consistent with
observations is produced due to catalytic effects as well.
Unfortunately, scenarios of stau NLSP
with a heavy gravitino LSP are most likely
untestable at the LHC, since $m_{\tilde{\tau}}\simge 1\,$TeV is preferred. 
This is different for scenarios with light
gravitinos. Here, either cosmologically important abundances of \li6 
(and possibly \be9) may be synthesized (for stau NLSP) or the \li7 problem
may be solved (for neutralino NLSP) for NLSPs light enough to be 
detectable at the
LHC. However, in such scenarios only a small fraction of the dark matter
is produced, with the remaining gravitino 
dark matter possibly produced during a 
reheating period after inflation.

The authors acknowledge useful discussions with Asimina Arvanitaki, 
Savas Dimopoulos, Peter Graham, Lawrence Hall, Masayasu Kamimura,
John March-Russell, Maxim Pospelov, and Frank Steffen.
This work was supported in part by ANR under contract
{\sc NT05-1\_43598/ANR-05-BLAN-0193-03}.

\bibliographystyle{h-physrev2}
\bibliography{biblio}

\begin{mcitethebibliography}{10}
\bibitem{remark0}
{Eq.(1) is a simplified form which encompasses the various NLSP types and
  2-body decays, such as stau NLSP decaying into $\tau \tilde{G}$, neutralino
  NLSP decaying into $\gamma \tilde{G}$, or into $Z \tilde{G}$. For instance,
  in the limit of pure bino-like neutralino and neglecting for simplicity the
  masses of all decay products except for the gravitino, the three cases above
  correspond to ($\kappa, n$) taking respectively the values $(1, 4)$, $(\simeq
  \rho\, cos^2 \theta_W , 3)$ and $(\simeq \rho\, sin^2 \theta_W , 4)$ where
  $\rho =(1 + 3 (m_{\tilde{G}}^2/M_{\rm NLSP}^2))$ and $\theta_W$ is the weak
  angle. We do not include here the higgs final state channel of the neutralino
  decays which is suppressed due to the almost pure bino content of the latter.
  The analyses presented in this paper rely however on the exact expressions
  for the lifetimes without any approximation.}\relax
\mciteBstWouldAddEndPunctfalse
\mciteSetBstMidEndSepPunct{\mcitedefaultmidpunct}
{}{\mcitedefaultseppunct}\relax
\EndOfBibitem
\bibitem{remark1}
{Note that whereas catalytic effects are mostly dependent on $Y_X = n_X/n_B$
  the X-particle-to-baryon ratio, hadronic effects are mostly dependent on
  $\Omega_Xh^2B_h \sim Y_X M_X B_h$, making it unpratical to show results
  dependent on either $\Omega_Xh^2B_h$ or $Y_X$.}\relax
\mciteBstWouldAddEndPunctfalse
\mciteSetBstMidEndSepPunct{\mcitedefaultmidpunct}
{}{\mcitedefaultseppunct}\relax
\EndOfBibitem
\end{mcitethebibliography}

\end{document}